\documentclass[
notitlepage,
bibnotes,
amsmath,amssymb,
aps,
10pt,
twocolumn,
prx,
]{revtex4-2}
\usepackage{amssymb,bm,bbold,bbm,amsmath}
\usepackage{xcolor}
\colorlet{mylinkcolor}{blue!66!black!80}

\newcommand{\e}{{\rm e}}

\newcommand{\blue}[1]{\textcolor{black}{#1}}

\usepackage{mathtools}
\usepackage[colorlinks=true,linkcolor=mylinkcolor,citecolor=mylinkcolor,filecolor=cyan,urlcolor=mylinkcolor,breaklinks=true]{hyperref}

 \usepackage{booktabs} 
\usepackage[utf8]{inputenc}
\usepackage{lmodern}
\usepackage{microtype}

\newcommand{\mtau}{\langle\tau\rangle}
\newcommand{\ttilde}{\tilde{t}}
\newcommand{\ntau}{\overline{\tau}_n}
\newcommand{\maxtau}{\tau_n^{+}}
\newcommand{\mintau}{\tau_n^{-}}
\newcommand{\C}{\mathcal{C}}
\newcommand{\tlambda}{\tilde{\lambda}}
\newcommand{\Ctt}{\mathcal{C}_{\tilde{t}}}

\newcommand{\bwk}{\bar{w}_k(p_0)}
\DeclareMathOperator{\sgn}{sgn}
\newcommand{\tp}{t^+_{\alpha_+}}
\newcommand{\tm}{t^-_{\alpha_-}}
\newcommand{\w}{w_{\tilde{t}}}
\newcommand{\y}{y_{\tilde{t}}}
\newcommand{\z}{z_{\tilde{t}}}
\newcommand{\Y}{Y_{\tilde{t}}}

\newcommand{\kplus}{{k_+}}
\renewcommand{\epsilon}{\varepsilon}
\newcommand{\napprox}{\not\approx}

\makeatletter
\newcommand*\rel@kern[1]{\kern#1\dimexpr\macc@kerna}
\newcommand*\widebar[1]{%
  \begingroup
  \def\mathaccent##1##2{%
    \rel@kern{0.8}%
    \overline{\rel@kern{-0.8}\macc@nucleus\rel@kern{0.2}}%
    \rel@kern{-0.2}%
  }%
  \macc@depth\@ne
  \let\math@bgroup\@empty \let\math@egroup\macc@set@skewchar
  \mathsurround\z@ \frozen@everymath{\mathgroup\macc@group\relax}%
  \macc@set@skewchar\relax
  \let\mathaccentV\macc@nested@a
  \macc@nested@a\relax111{#1}%
  \endgroup
}
\makeatother

\begin{document}
\title{Controlling Uncertainty of Empirical First-Passage Times in the
  Small-Sample Regime}
\author{Rick Bebon}
\author{Alja\v{z} Godec}%
\email{agodec@mpinat.mpg.de}
\affiliation{Mathematical bioPhysics Group, Max Planck Institute for Multidisciplinary Sciences, 37077 Göttingen, Germany}
\begin{abstract}
We derive general  bounds on the probability that the \emph{empirical}
first-passage time $\overline{\tau}_n\equiv \sum_{i=1}^n\tau_i/n$ of a
reversible ergodic Markov
	process
inferred from a sample of
    $n$ independent realizations deviates from the true mean
first-passage time by more than any given
    amount in either direction. We construct
    non-asymptotic confidence intervals that hold in the elusive
small-sample regime and thus fill the gap between asymptotic
methods and the Bayesian approach that is known to be sensitive to prior belief and tends to
underestimate uncertainty in the small-sample setting. {We prove
sharp bounds on extreme first-passage times that control uncertainty
even in cases
where the mean alone does not sufficiently characterize the statistics.}
Our
concentration-of-measure-based results allow for model-free
error control and reliable error estimation in kinetic inference, and
are thus
important for the
analysis of experimental and simulation data in the presence of
limited sampling. 
\end{abstract}

\maketitle
\blue{The first-passage time
$\tau$ denotes the time a random 
process 
reaches a
threshold $a$, 
referred to as 
 ``target'', for the first time.~First-passage phenomena~\cite{redner2001guide, metzler2014first,Zhang2016,iyer2016first} 
are ubiquitous; 
they 
quantify the kinetics of physical, 
chemical,
and biological~\cite{Kramers,Szabo,Haenggi,Redner,Grebenkov_2018,Grebenkov_2018_2,Grebenkov_PRL1,TrendelkampSchroer2016,Satya_PRL,Boccardo2022,Chupeau2019,Wales_2020,Tolya,Goychuk2019,Bressloff,Roldn2016,Ghusinga2017,Rijal2022,Parmar2015,Frey2019,Lloyd2001,Hufnagel2004,Olivier_RMP,Bebon2022,Erdmann2004_2,Thirumalai,Blom2021} 
processes from the low-copy~\cite{Berg,Andy,Holcman,Olivier_Chromatin,Marklund,Bauer,Olivier_NC,Bnichou2014,Godec_PRX,Newby_PRL}
to the ``fastest 
encounter''~\cite{Redner_2,Redner_3,Hartich_2018,Hartich_2019,Hartich_Book,Schuss_2019,Lawley_1,Lawley_2,Lawley_4}
limit, 
characterize 
persistence properties~\cite{Dougherty2002,Constantin2003,Dougherty2005,Merikoski2003,Constantin2004,Godrche2009, Bray_2013}, 
search processes
~\cite{condamin2007first,condamin2005first,condamin2007random,meyer2011universality,chevalier2010first,MejaMonasterio2011,Oshanin2011,Mattos2012},
and fluctuations of path-observables in stochastic 
thermodynamics~\cite{currents,Singh2019,decision,stopping,Falasco2020,Neri2022,Garrahan2017,Hiura2021}.~First-passage 
ideas are 
tied 
to the statistics of extremes~\cite{Kac,Schehr,Satya_records,Hartich_JPA},
and were 
extended to quantum systems~\cite{Eli_QM1,Eli_QM},
additive functionals~\cite{Kearney2005,Kearney2007,Kearney2014,Kearney2021,Majumdar2021,Singh2022}, 
intermittent targets~\cite{MercadoVsquez2019,Kumar2021,Spouge1996,Scher2021},
active particles~\cite{Woillez2019,Mori2020,DiTrapani2023},
non-Markovian dynamics~\cite{Hnggi1983,Hanggi1985,Gurin2016,Meyer2021}, 
and resetting processes~\cite{Evans2011,Kusmierz2014,Reuveni2016,Pal2017,Pal2019,Evans2020,Besga2020,TalFriedman2020,DeBruyne2020,DeBruyne2022}.
}
\\
\indent Whereas theoretical studies focus on predicting first-passage 
statistics, practical applications
typically aim at
inferring kinetic rates---
inverse mean first-passage times---
from experiments
\cite{Pande,Joerg_2020,RNA_2017,Single_cell_snapshot,folding_genes_HIV,Tolya}
or simulations 
\cite{all_atom_JACS,MD_Netz,MD_nucleation,MD_water_pores,Edgar_MD,OUP,Wales_2020,Bert}.~The
inference of \emph{empirical first-passage times} $\overline{\tau}_n\equiv\sum_{i=1}^n\tau_i/n$
from data
is, however, challenging 
because usually only
a small number of realizations $n$ (typically 1-10~\cite{lindorff2011fast,Adelman2013,Bert,Mostofian2019,Mehra2019,Militaru2021},
sometimes up to 100~\cite{Rondin2017}) are available, which gives rise to
large uncertainties and non-Gaussian errors.~Sub-sampling issues are
especially detrimental in the case of  broadly
distributed~\cite{Wales_2020,Sharpe2020,Donovan2013} and high-dimensional
data~\cite{folding_genes_HIV}.~Moreover, first-passage times are generically
\emph{not} exponentially distributed  
\cite{Sabelko1999,Ensign2009,stigler2011complex,Olivier_NC,Godec_PRX,Hartich_2018,Grebenkov_2018,Grebenkov_2018_2,Hartich_2019,Berezhkovskii2019,Nayak2020,Wales2022}, 
which further complicates 
quantification of uncertainty.~A systematic understanding of
statistical
deviations of the empirical 
from the true mean
first-passage time 
(see Fig.~\ref{fig:fig1}a), especially in the
small-sample $n\lesssim100$ regime, remains elusive. 
\begin{figure}[ht!]
\centering
\includegraphics[width=0.85\columnwidth]{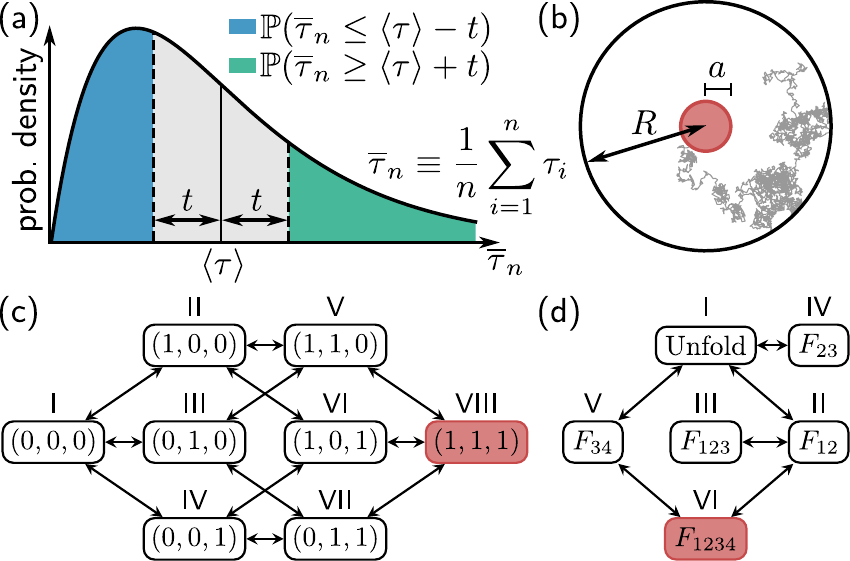}
\caption{Deviations of empirical first-passage times from the true mean and model systems.
(a) Schematic probability density of empirical first-passage time $\ntau$ inferred from a sample of $n$ realizations of an ergodic 
reversible Markov process.~The tail probability that the estimate $\ntau$ deviates from the true mean $\mtau$ by more or 
equal than $t$ upwards $\mathbbm{P}(\ntau \geq \mtau + t)$ 
or downwards $\mathbbm{P}(\ntau \leq \mtau - t)$ is shown in green and
blue, respectively.~(b)~Brownian molecular search process in a
$d$-dimensional domain (here $d=2$) with outer radius $R$ and target
radius $a$.
Discrete-state Markov jump models of protein folding
for (c) a toy protein and (d) experimentally inferred model of
calmodulin~\cite{stigler2011complex}.
Transitions between states obey detailed balance and absorbing targets are colored red.
}	
\label{fig:fig1}
\end{figure}

Computer simulations often especially suffer from insufficient sampling, which
leads 
to substantial
errors in inferred rates~\cite{Singhal2005,bowman2013introduction,grossfield2009quantifying,grossfield2018best}
and, in the worst case, 
erroneous
conclusions (see discussion in \cite{Bert,Knapp2018}). Even
extensive computing resources may result in only a few
independent estimates spread over many orders of magnitude, rendering
uncertainty quantification challenging and \emph{not} amenable to
standard error analysis
~\cite{Mostofian2019}.\\    
\indent Constructing \emph{reliable} confidence 
intervals is a fundamental challenge in statistical inference, and many 
prevalent methods only
hold
when $n\to\infty$.
The applicability
of such asymptotic results in a finite-sample setting is, by definition,
problematic. 
In particular,
Central-Limit- and bootstrapping-based methods 
\footnote{Resampling methods like bootstrapping assume the 
data to be representative of the inferred statistic, which is not
necessarily the case for small $n$, possibly even when $n$ is large but
finite for broad distributions.} may easily underestimate the
uncertainty for small $n$ and
fail to guarantee coverage of the confidence 
level~\cite{Davison1997,Shao2003,Imbens2008,Putter2011,hogg1995introduction,schenker1985qualms,Mostofian2019}.\\    
\indent {
Conversely, Bayesian methods
(e.g.~\cite{gelman1995bayesian}), 
despite not relying on asymptotic arguments, must be
treated with care, 
as 
estimates and their uncertainties
are 
sensitive to, dependent on, and potentially
biased by, the specification of the prior
distribution, \emph{especially}
in the small-sample setting~\cite{Adelman2013,Smid2019}
(see~\cite{Ensign2009,Bacallado2009,bowman2013introduction,Prinz2011_2,TrendelkampSchroer2015}
for kinetic inference).
Moreover, prior-dependent uncertainty estimates seem to 
remain, even in the asymptotic limit, an elusive problem
(see extended discussion in \cite{Note5}).}\\
\indent There is thus a pressing need for understanding fluctuations
of inferred empirical first-passage times, a rigorous error control, and
reliable \emph{non-asymptotic} error estimation 
in the small-sample regime. These are fundamental {unsolved }problems
of statistical kinetics and are essential 
for the analysis of
experimental and simulation data.\\    
\indent Here, we present general bounds on fluctuations of \emph{empirical 
first-passage times} that allow a rigorous uncertainty quantification
(e.g.~using confidence intervals with guaranteed coverage
probabilities for all sample sizes) under minimal assumptions. 
We prove 
non-asymptotic lower ($\mathcal{L}$) and upper ($\mathcal{U}$)  bounds on 
the deviation probability $\mathbbm{P}(\ntau\geq\mtau +t)$ and $\mathbbm{P}(\ntau\leq\mtau -t)$
(see Fig.~\ref{fig:fig1}a), i.e., 
the probability that the empirical first-passage time inferred
from a sample of $n\geq 1$ 
realizations of an ergodic reversible Markov process,
$\overline{\tau}_n$, 
deviates from the true mean $\langle \tau\rangle$ by more than
$t$ in either direction, 
\begin{equation}
\mathcal{L}^\pm_n(t) \leq 
\mathbbm{P}(\pm[\ntau-\langle \tau\rangle]\ge t)\le
\mathcal{U}_n^{\pm}(t)\quad \forall t\ge 0,
  \label{prob}
\end{equation}
the upper bounds $\mathcal{U}_n^{\pm}(t)$ corresponding to so-called
\emph{concentration inequalities}
\cite{boucheron2013concentration}.~The most conservative version of
the derived upper bounds is \emph{independent of any details about the
underlying dynamics}.
{
We use the 
bounds $\mathcal{U}^{\pm}_n(t)$ to quantify the uncertainty of the
inferred sample mean $\ntau$ \emph{in a general setting and under minimal assumptions},
for all $n\geq 1$. 
We further derive general lower ($\underline{\mathcal{M}}$) and upper ($\widebar{\mathcal{M}}$) bounds
on 
the expected minimum $\tau_n^-=\min_{i\in[1,n]}\tau_i$ and maximum
$\tau_n^+=\max_{i\in[1,n]}\tau_i$ of 
$n$ 
realizations, i.e.\
\begin{equation}
\underline{\mathcal{M}}_n^\pm \leq \langle \tau^\pm_n -\mtau \rangle \leq \widebar{\mathcal{M}}_n^\pm\quad \forall n\ge 1,
\end{equation}
controlling the uncertainty of first-passage times
 even when multiple
  time-scales are involved, rendering $\mtau$ an \emph{a priori}
  insufficient statistic.} 
%
The validity and sharpness of 
bounds are
demonstrated by means of spatially confined Brownian 
search processes in
dimensions 1 and 3 (Fig.~\ref{fig:fig1}b), and discrete-state
Markov jump models of protein
folding for a toy protein ~\cite{Prinz2011,bowman2013introduction,Olsson2017,Hartich_2019}
(Fig.~\ref{fig:fig1}c)
and the experimentally inferred model of
calmodulin~\cite{stigler2011complex} (Fig.~\ref{fig:fig1}d). 
We conclude with a discussion of the practical implications of the results and further research directions.   

\emph{Setup.---}We consider time-homogeneous Markov
processes $x_t$ on a continuous or discrete state-space $\Omega$ with
generator
$\hat{L}$ corresponding to a Markov rate-matrix or an effectively
one-dimensional Fokker-Planck
operator. 
Let the transition probability density to find $x_t$ at $x$ at time $t$ given that
it evolved from $x_0$ be $p_t(x|x_0)\equiv
\mathrm{e}^{\hat{L}t}\delta_{x_0}(x)$ where $\delta_{x_0}(x)$ denotes
the Dirac or Kronecker delta for continuous and discrete state-spaces,
respectively. We assume the process to be ergodic
$\lim_{t\to\infty}p_t(x|x_0)=p_{\rm eq}(x)$, where $p_{\rm
  eq}(x)\equiv\mathrm{e}^{-\varphi(x)}$ denotes
the equilibrium probability density and $\varphi(x)$ the 
generalized potential in units of thermal energy $k_{\rm B}T$~\cite{Pavliotis}.
We assume that $\hat{L}$ 
obeys detailed balance 
\footnote{
The operator $\mathrm{e}^{\varphi(x)/2}\hat{L}\mathrm{e}^{-\varphi(x)/2}$ is
self-adjoint.} 
and is either 
(i) bounded,
(ii) $\Omega$ is finite with reflecting boundary $\partial \Omega$, 
or (iii) $\Omega$ is infinite but $\varphi(x)$ 
sufficiently
confining (see \footnote{Precisely, we require that $\varphi(x)$ 
satisfies the
Poincar\'e inequality,
i.e.\ $\lim_{|x|\to\infty}(|\nabla\varphi(x)|^2/2-\nabla^2\varphi(x))=\infty$.}).~Each
of the conditions (i)-(iii) ensures that the 
spectrum
of $\hat{L}$ is discrete \footnote{The relaxation eigenvalue
problem reads $-\hat{L}\Psi_k(x)=\nu_k\Psi_k(x)$ with $\nu_0=0$
and $\nu_{k\ge 1}>0$~\cite{Pavliotis}}.

We are interested in the first-passage time 
to a target $a$ when
$x_{t=0}$ is drawn from a density $p_0(x)$
\begin{equation}
\tau=\inf_t [\,t\, |x_t= a, \, p_0(x_0)],
  \label{fpt}
\end{equation}
and focus on 
{the 
  setting
$p_0(x)=\tilde{p}_{\rm eq}(x)$, since $x_0$ usually cannot be
  controlled  experimentally
(see e.g.,~\cite{Bnichou2014, Olivier_NC, Mattos2012, chevalier2010first, meyer2011universality, bowman2013introduction,noe2019boltzmann,Braun2019}}),
and the tilde denotes that the absorbing state is
excluded \blue{(see Appendix Ia for details).} 
For completeness we also provide
results for general initial conditions
$p_0(x)$ 
\blue{in Appendix Ib and} \footnote{See Supplemental Material at [...] for further
  details, mathematical proofs, and generalizations to arbitrary
  initial conditions $p_0(x)$, as well as Refs.~\cite{SM_burden2015numerical,SM_Barkai2014,SM_Pitman1975,SM_Siegert1951,SM_Seifert2019,SM_cowan1998statistical}}
that
require more precise conditions on
$\varphi(x)$. 
The probability density of $\tau$ 
for such 
processes 
has the generic form~\cite{Hartich_2018,Hartich_2019} 
\begin{equation}
\wp_a(t|x_0)=\sum_{k>0}\mu_kw_k^{x_0}\e^{-\mu_k t},
  \label{FPT_den}
\end{equation}
with first-passage rates $\mu_k>0$ and
(not necessarily positive) spectral weights $w_k^{x_0}$ normalized according
to $\sum_{k>0}w_k^{x_0}=1$ and $w_1^{x_0}>0$.
The $m$-th moment of $\tau$ is 
given by 
$\langle \tau^m\rangle= m! \sum_{k>0}w^{x_0}_k/\mu_k^m$ 
and the 
\emph{survival probability} reads
$\mathbbm{P}(\tau>t)\equiv S_a(t|x_0)=\sum_{k>0}w_k^{x_0}\e^{-\mu_kt}$.
If $x_0$ is drawn from the equilibrium density, $\tilde{p}_{\rm eq}(x)$,
we have 
$\wp_a(t|\tilde{p}_{\rm eq})\equiv
\int_{\Omega\setminus a}\wp_a(t|x_0)\tilde{p}_{\rm
  eq}(x_0)dx_0$~\footnote{When $\Omega$ is discrete the integral is
replaced by a sum over states excluding the target.}
which renders all
weights non-negative, ${w_k\,\equiv\,\int_{\Omega\setminus a}w_k^{x_0}\tilde{p}_{\rm eq}dx_0\ge 0}$~(see proof in \cite{Note5}).~We henceforth abbreviate ${S_a(t|\tilde{p}_{\rm eq})\equiv S_a(t)}$.

To exemplify the need for uncertainty bounds 
in Eq.~\eqref{prob} 
we show in Fig.~\ref{fig:fig2}a-d that the probability that
${\ntau-\langle\tau\rangle}$ lies within a desired range of say $\pm$ 10\% of the
\emph{longest} first-passage time scale
$\mu_1^{-1}$, $\mathbbm{P}(\mu_1[\ntau-\langle\tau\rangle]\in
[-0.1,0.1])$ is low even for $n\approx 50$ for all models in Fig.~\ref{fig:fig1}b-d. 
\blue{This inherent \emph{intrinsic noise-floor}
of the inferred observable $\ntau$ for any $n$ is embodied in, and can be explicitly demonstrated by, the existence of lower bounds $\mathcal{L}^\pm_n$ (see Appendix II).}
     
\begin{figure*}
\centering
\includegraphics[]{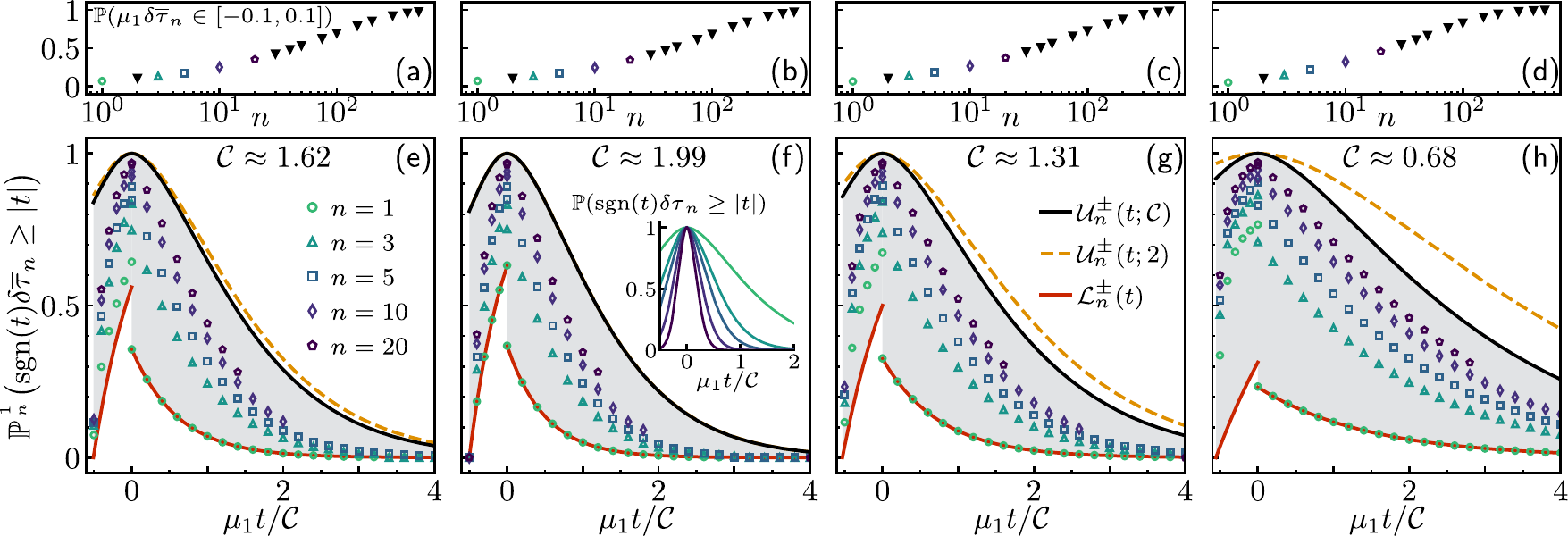}
\caption{
Deviation probabilities and corresponding bounds for a spatially confined Brownian search process in (a,e) $d=1$ and (b,f) $d=3$ dimensions,
and Markov-jump models of protein folding for 
(c,g) the experimentally inferred model of calmodulin and (d,h) the toy protein.
(a-d) Probability that $\delta\ntau = \ntau-\mtau$ lies within a range of 
$\pm 10\%$ of the longest time-scale $1/\mu_1$, $\mathbbm{P}(\mu_1\delta\ntau\in[-0.1,0.1])$,
as a function of $n$ determined from the statistics of $\ntau$ for different fixed $n$ for all model systems.
(e-h) Scaled probabilities $\mathbbm{P}^{1/n}(\text{sgn}(t)\delta\ntau \geq |t|)$ that the sample mean $\ntau$
inferred from $n$ realizations deviates from $\mtau$ by more than $t$ in either direction.
Right tail areas are shown for $t>0$ and left for $t<0$, respectively.
Lower $\mathcal{L}^\pm_n(t)$ and upper $\mathcal{U}^\pm_n(t;\C)$ bounds are depicted
as red and black lines, respectively, and the model-free upper bound
$\mathcal{U}^\pm_n(t;2)$ as the dashed yellow line.
Symbols denote corresponding scaled empirical deviation probabilities as a function of $t$ and are sampled for different $n$.}	
\label{fig:fig2}
\end{figure*}

\emph{Cram\'{e}r-Chernoff bounds.---}\blue{To tackle this issue we now prove 
upper bounds using the Cram\'er-Chernoff approach.}
Let
$\delta\ntau\equiv|\ntau-\mtau|$ and $\lambda\in\mathbbm{R}^+$. We start with the 
inequality
$\mathrm{e}^{\lambda t}\mathbbm{1}_{\delta\ntau\ge t}\le\mathrm{e}^{\lambda\delta\ntau}$,
where $\mathbbm{1}_{b}$ is the indicator function of the set
$b$. Taking the expectation 
yields
$\mathbbm{P}(\delta\ntau\ge t)\le \mathrm{e}^{-\lambda t}\langle
\mathrm{e}^{\lambda\delta\ntau}\rangle\equiv\mathrm{e}^{-\lambda
  t+\psi_{\delta\ntau}(\lambda)}$, where we defined the
cumulant generating function of $\delta\ntau$ as $\psi_{\delta\ntau}(\lambda)\equiv\ln\langle
\mathrm{e}^{\lambda\delta\ntau}\rangle$.~Note that
$\tau_i$ are
statistically independent.~The bound can be
optimized \footnote{$\psi_{\delta\tau}(\lambda)$ and $\phi_{\delta\tau}(\lambda;\C)$ are differentiable, convex, non-negative, and
non-decreasing functions in $\lambda$. The optimization is thus carried out as $\psi^\ast_{\delta\tau}(t)=\psi_{\delta\tau}(\lambda^\dagger),$ where
$\lambda^\dagger$ solves $\psi'_{\delta\tau}(\lambda^\dagger)=t$. The computation for $\phi_{\delta\tau}(\lambda;\C)$ follows analogously.}
to find Chernoff's inequality
{
$\mathbbm{P}(\delta\ntau\ge
t)\le\mathrm{e}^{-n\psi^*_{\delta\tau}(t)}$} 
where $\psi^\ast_{\delta\tau}(t)$
is the Cram\'er transform of $\psi_{\delta\tau}(\lambda)$ 
\cite{boucheron2013concentration},
\begin{equation}
\psi^\ast_{\delta\tau}(t)\equiv\sup_\lambda(\lambda
  t-\psi_{\delta\tau}(\lambda)),
\label{Cramer_tr}
\end{equation}
with $\delta\tau\equiv\delta\overline{\tau}_1$. On
the interval
$\lambda\in[0,\mu_1)$ we have the following bounds on 
  $\psi_{\delta\tau}(\lambda)$ 
 (see proof in \cite{Note5})
\begin{align}
  \psi_{\delta\tau}(\lambda)\le \phi_{\delta\tau}(\lambda;\C)\equiv
 \begin{dcases} 
  \frac{\lambda^2}{2\mu_1^2}\frac{\C}{1-\lambda/\mu_1}&\tau\ge\mtau\\
  \frac{\lambda^2}{2\mu_1^2}\frac{\C}{1-(\lambda/\mu_1)^2}&\tau<\mtau,
\end{dcases}   
\label{bounds}
\end{align}
which are non-negative, convex, and increasing
on $\lambda\in[0,\mu_1)$, and
we introduced $\C\equiv \mu_1^2 \langle\tau^2\rangle$.
\blue{Note that general initial conditions $p_0(x_0)$ are accounted
  for by simply replacing $\langle\tau^2\rangle$ in $\C$~(see Appendix Ib and~\cite{Note5} for details).}
The bound~\eqref{bounds}
further implies $\psi_{\delta\tau}^*(t)\geq\phi_{\delta\tau}^*(t;\C)$, $\forall t\geq 0$,
 and may thus be optimized \cite{Note7}
  to obtain the inequalities 
  announced in Eq.~\eqref{prob}
  via Chernoff's inequality:
\begin{align}
\mathcal{U}_n^+(t;\C)
&=\exp\left(-n\C  h_+\left(\mu_1 t/\C\right)\right)
\quad 0 \leq t \leq \infty
\nonumber \\
\mathcal{U}_n^-(t;\C)
&=\exp\left(-n\C h_-\left(\mu_1 t /\C\right)\right)
\quad 0 \leq t \leq \mtau
\label{eq:conc_n1}
\end{align}
where we defined 
the 
functions
\begin{align}
h_+(u)&\equiv 1+u-\sqrt{1+2u}
\\
h_-(u)&\equiv \Lambda(u)u-\frac{1}{2}\frac{\Lambda(u)^2}{1-\Lambda(u)^2}
\end{align}
with $\Lambda(u)\equiv
\frac{1}{2}\left[g(u)-\sqrt{4+2/g(u)u -g(u)^2}\right]$
and
\begin{equation}
\!g(u)\equiv\frac{2}{\sqrt{3}}\left\{1+2\cosh\left[\frac{1}{3}{\rm arcosh}\left(1+\frac{3^3}{2^7u^2}\right)\right]\right\}^{1/2}\!\!\!\!\!.
\label{last}
\end{equation}
{These results control the deviations of the inferred $\ntau$
  from $\mtau$  for all $n$.  
  Note that 
  in general whenever $\mu_1\mtau\napprox 1$
  (e.g. in ``compact'' search processes
  \cite{Olivier_NC,Bnichou2014}),
  $\mtau$ is not necessarily representative due to the
  multiple relevant
  time-scales
  \cite{MejaMonasterio2011,Oshanin2011,Mattos2012}.~However, as we
  show below, $\mtau$ sharply bounds extreme first-passage times.
  Thus, inferring $\ntau$ 
  provides insight about first-passage
  statistics even when $\mtau$ is \emph{a priori} not a sufficient statistic. 
}

\blue{Concerning practical applications,}
deviations are readily expressed 
\blue{relative} to
the \emph{longest natural} time-scale $1/\mu_1$ that does \emph{not} need to be
known. 
That is, 
errors \blue{in units of $1/\mu_1$}, 
$\mu_1(\ntau-\mtau)$,
are naturally parameterized by the \emph{dimensionless}
variable $\mu_1t$. 
Remarkably, details about the underlying dynamics \blue{then} only enter the 
bounds~\eqref{eq:conc_n1} via a system-dependent
constant $\C$ that, however, 
can be bounded.~In particular, for
equilibrium initial conditions we have $0\leq  2w_1 \leq \C \leq 2$ (see \cite{Note5}).
Since 
$\phi_{\delta\tau}$  is monotonically increasing with $\C\in(0,2]$,
we have $\phi_{\delta\tau}(\lambda;\C)\leq\phi_{\delta\tau}(\lambda;2)$
which implies $\phi^*_{\delta\tau}(t;\C)\geq\phi^*_{\delta\tau}(t;2)$.
Thus, we find
the \emph{model-free} bounds \blue{by setting $\C\equiv 2$}
\begin{equation}
\mathcal{U}_n^\pm(t;\C)\le \mathcal{U}_n^\pm(t;2)\equiv \mathcal{U}_n^\pm(t)
\label{eq:c_bound}
\end{equation}
requiring no information about the
system.~The \emph{non-asymptotic} bounds on deviation probabilities
of $\ntau$ 
in Eqs.~\eqref{eq:conc_n1} and \eqref{eq:c_bound} are our first main
result.
\blue{Their most general and conservative version states
that for all sample-sizes $n$, the probability of observing a relative error
larger than a specified value $\mu_1 t$ (e.g., $\mu_1 t=0.1$ or $10\%$), 
$\mathbbm{P}(\mu_1[\ntau-\mtau]\geq \mu_1 t)$, 
is lower than the corresponding upper bound $\mathcal{U}_{n}^\pm(\mu_1 t)$, which solely depends on $\mu_1 t$.}

Lastly, as $n\to\infty$, $\mathcal{U}^\pm_n$ is substantial only for
$\mu_1 t /\C\ll 1$
and the tails become symmetric and sub-Gaussian~\cite{boucheron2013concentration}, $h_+(u)=u^2/2-\mathcal{O}(u^3)$ and 
$h_-(u)=u^2/2-\mathcal{O}(u^4)$ (see \cite{Note5}).
Notably, concentration
inequalities were recently derived
for time-averages of ``classical'' 
\cite{Lezaud_1,Lezaud_2,Wu} (see \footnote{Ref.~\cite{Lezaud_2}
contains an error;~the Proof of Lemma 2.3 is only valid in the regime
$r<\lambda_1/3||f||_\infty$, but the Lemma may be shown to hold
in the claimed regime \cite{Santiago}.}), and 
quantum \cite{Garrahan_QM} Markov
processes,  as well as inverse thermodynamic uncertainty
relations \cite{Garrahan_ITUR}.

\emph{Illustration of {deviation} bounds.---}The upper $\mathcal{U}^\pm_n(t)$ 
and lower $\mathcal{L}^\pm_n(t)$
bounds on $\mathbbm{P}(\pm[\ntau-\mtau]\geq t)$ in
Eqs.~\eqref{eq:conc_n1} and \eqref{lower},
respectively, are examplified in Fig.~\ref{fig:fig2}e-h (see black and
red lines) for the model
systems shown in Fig.~\ref{fig:fig1}b-d.
Note that to illustrate all bounds 
we formally let $t\to -t$
for the left tails $\mathcal{L}^-_n(t)$ and $\mathcal{U}^-_n(t)$, such
that their support is 
on $[-\mtau, \infty)$. Deviation probabilities are in turn expressed
  as $\mathbbm{P}(\text{sgn}(t)\delta\ntau \geq |t|)$ where $\text{sgn}(x)$
  denotes the
  signum function
  and $\delta\ntau = \ntau -\mtau$.
  
To assess the quality of our bounds 
we scale probabilities $\mathbbm{P}^{1/n}$ 
such that 
$\mathcal{L}^\pm_n(t)$ and
$\mathcal{U}^\pm_n(t)$
collapse onto a master curve for all $n$ 
(see inset Fig.~\ref{fig:fig2}f).
Symbols denote empirical deviation probabilities obtained by sampling $\ntau$ for different 
$n$ (see \cite{Note5} for details), 
which approach the upper bound as $n$ increases.
For $n=1$ right-tail deviations are close to
$\mathcal{L}^+_1(t)$ even for $w_1\leq 1$ \footnote{However, $w_1$ can get arbitrary close to 0 in principle,
rendering the lower bound trivial.}.
As expected the model-free bound $\mathcal{U}^\pm_n(t;2)$ (yellow) holds universally but is generally 
more conservative. 
{Notably, it remains remarkably good even for $\C\gtrsim 1.3$ (Fig.~\ref{fig:fig2}e-g).}\\ 
%
\indent \emph{Uncertainty quantification.---}The 
bounds~\eqref{eq:conc_n1} and~\eqref{eq:c_bound}
provide the elusive 
systematic framework to \emph{rigorously quantify the uncertainty} of the
estimate $\ntau$ for any, and especially for small, sample sizes.
In particular, they allow the construction of
``with high probability'' 
guarantees such as 
confidence intervals, which---unlike traditional confidence intervals
in statistics---are \emph{not only} asymptotically correct but hold
for any $n$.
Concentration-based guarantees do further \emph{not} require
specifying a prior belief as in the Bayesian context.
Setting $\mathcal{U}_n^\pm( t^\pm_{\alpha_\pm};\C)=\alpha_\pm$  for chosen acceptable
left- and right-tail error probabilities $\alpha_\pm$ (with $\alpha_++\alpha_-<1$), we get an
implicit definition of the confidence interval $[-t^-_{\alpha_-},
  t^+_{\alpha_+}]$ at confidence level (or ``coverage probability'')
$1-(\alpha_- + \alpha_+)$ in the form
\begin{align}
\mathbbm{P}(-t^-_{\alpha_-} \leq \delta\ntau \leq t^+_{\alpha_+}) \geq
1-\alpha_- - \alpha_+\equiv 1-\alpha,
\label{eq:conf_int}
\end{align}
stating that 
with probability of \emph{at least} $1-\alpha$
the sample mean $\ntau$ lies within $[\mtau-t^-_{\alpha_-},
  \mtau+t^+_{\alpha_+}]$. 
Confidence intervals are closely related to, and can be used for, 
statistical significance
tests~\cite{wackerly2014mathematical,meeker2017statistical}, and
beyond that provide quantitative bounds on statistical
uncertainty.
Two-sided 
central confidence intervals for $\delta\ntau$ as a function of $n$ for a confidence
level of $\alpha=0.1$ and models systems in Fig.~\ref{fig:fig1}b-d are
shown (rescaled to a master scaling) in Fig.~\ref{fig:fig3}a (for a detailed discussion see \cite{Note5}).

In
particular, we may now also answer the practical question: \emph{How
many realizations are required to achieve a desired
accuracy with a specified probability?}
To ensure with probability of at least $1-\alpha$ that 
$\delta\overline{\tau}_{n^*}\in[-t^-_{\alpha_-},t^+_{\alpha_+}]$ one needs 
\blue{a minimal sample size}
$n^*$
defined via
\begin{equation}
\mathcal{U}_{n^\ast}^+( t^+_{\alpha_+};\C)+\mathcal{U}_{n^\ast}^-( t^-_{\alpha_-};\C)=\alpha.
\label{implicit}
\end{equation}  
The
number of samples $n^\ast$ required to \emph{guarantee} that $\overline{\tau}_{n^*}$
falls within a symmetric interval of length $\Delta t=0.2/\mu_1$, 
(i.e.~$\overline{\tau}_{n^*}\in[\mtau-0.1/\mu_1, \mtau+0.1/\mu_1]$) with
probability of \emph{at least} $1-\alpha$ 
is shown in Fig.~\ref{fig:fig3}b
for several values of $\C$ (intersections with the dashed 
line yield $n^\ast$ guaranteeing a coverage of \emph{at least}
90\%).~Fig.~\ref{fig:fig3}c depicts the complementary symmetric interval $\Delta
t$ covering the range of $\delta \ntau$ for a given $n$ with
probability of \emph{at least} 90\%. Note that hundreds to
thousands of samples may be required to ensure an accuracy of $\pm 0.1/\mu_1$
with a 90\% confidence, which is seemingly not met in experiments
\cite{lindorff2011fast,Adelman2013,Bert,Mostofian2019,Mehra2019,Militaru2021,Rondin2017}.\\
\indent Eqs.~(\ref{eq:conf_int}-\ref{implicit}) can be solved for $t^\pm_{\alpha_\pm}$ and
$n^\ast$, respectively, using standard root-finding
methods (see \cite{Note5}) and constitute our second main result.~They
allow for rigorous error control in kinetic inference in the small-$n$
regime.
\begin{figure}[hbtp]
\centering
\includegraphics[width=\columnwidth]{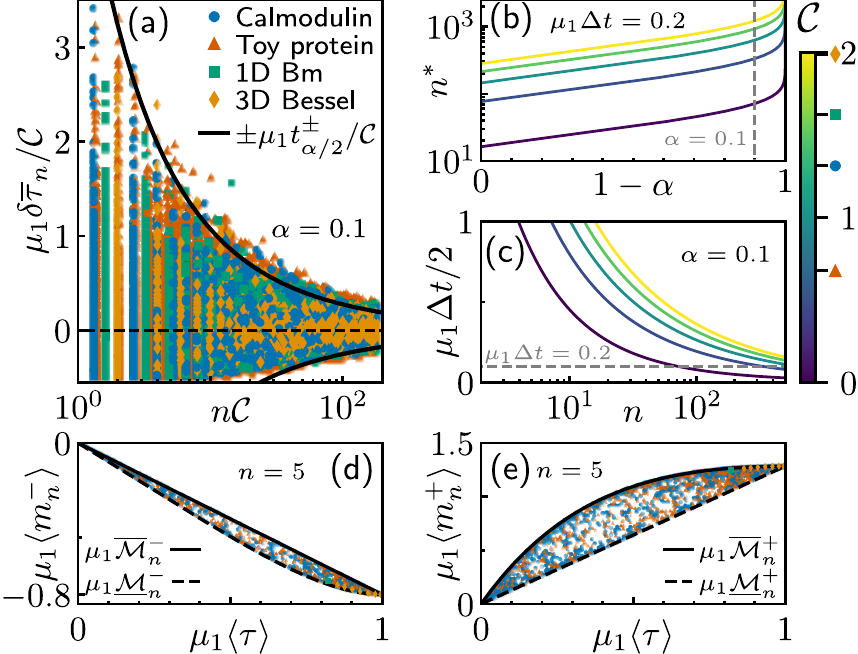}
\caption{Non-asymptotic uncertainty quantification of the sample mean $\ntau$
{and extreme deviations from $\mtau$.}
(a) Relative error $\mu_1\delta\ntau=\mu_1(\ntau-\mtau)$ (symbols) 
obtained from sampling of $\ntau$ for different $n$ and model systems. 
The central confidence interval $[-\mu_1 t^-_{\alpha/2}, \mu_1 t^+_{\alpha/2}]$ 
with $\alpha=0.1$ in black.
(b) Number of samples $n^*$ to ensure that 
the  error $\mu_1 \delta\overline{\tau}_{n^*}$ falls in the interval $[-0.1,0.1]$ of length
$\Delta t=0.2/\mu_1$
with probability of at least $1-\alpha$ for several values of $\C$.
(c)~Symmetric confidence interval $[-\mu_1 \Delta t/2, \mu_1 \Delta t/2]$ (upper limit shown) 
for $\alpha=0.1$ as a function of $n$ for different $\C$.
{Average extreme deviation from the mean $\langle m_n^\pm\rangle\equiv \langle\tau^\pm_n-\mtau \rangle$ (symbols) 
and bounds (lines) for the minimum (d) and maximum (e) 
of $n=5$ realizations for different models and values of $\mu_1\mtau$.}
}	
\label{fig:fig3}
\end{figure}
%
Using Eq.~\eqref{eq:c_bound} we can further construct
system-independent but more conservative
\emph{universal confidence intervals} (see yellow line in Fig.~\ref{fig:fig3}b,c).
Interestingly, even when $\C\approx 1$ the universal bound
remains reasonably tight, only for $\C\ll 1$ differences become substantial.\\ 
{\indent \emph{Bounding extreme deviations.---}Finally, we show that $\mtau$ controls
the range of inferred $\tau_i$ in any sample
of $n$ independent realizations, i.e., it \emph{sharply} bounds 
the average minimum $\tau_n^-$ and maximum $\tau_n^+$
deviations from the mean
$m^\pm_n\equiv\tau^\pm_n-\mtau$.
As our third main results we prove \blue{(see sketch of proof and
saturation conditions in Appendix~III and \cite{Note5} for details)} two-sided bounds
$\underline{\mathcal{M}}_n^{\pm}\leq \langle m^\pm_n\rangle \leq \widebar{\mathcal{M}}_n^{\pm}$,
where
\begin{align}
\label{eq:MaxMinBounds}
&\!\!\underline{\mathcal{M}}_n^{+}
\equiv 
\mtau \sum_{k=2}^n \frac{1}{k},
\,\,\,\,
\widebar{\mathcal{M}}_n^{+}
\!\equiv\!
\sum_{k=1}^n\!\binom{n}{k} (-1)^{k+1}\frac{(\mu_1\mtau)^k}{\mu_1 k}-\mtau
\\
&\!\!\underline{\mathcal{M}}_n^{-}
\equiv \mtau\!\left[n^{-1}(\mu_1\mtau)^{n-1} -1 \right],
\,\,\,\,\,\,
\widebar{\mathcal{M}}^-_n \equiv \mtau\!\left[n^{-1}-1\right].
\nonumber
\end{align}
\blue{Remarkably, the tight bounds on $\mu_1\langle m^\pm_n\rangle$ are
completely
  specified by the dimensionless parameter $\mu_1\mtau$.}
This motivates the inference of $\ntau$ in a general setting, even when $\mtau$ is \emph{a priori} not representative, since via the bounds~\eqref{eq:MaxMinBounds} (shown in
Fig.~\ref{fig:fig3}d-e)
$\mtau$ 
controls the range of estimates.
The bounds \eqref{eq:MaxMinBounds} provide insight
about the \emph{slowest} and \emph{fastest} out of $n$ first-passage times, 
and are thus relevant in the ``few encounter'' limit~\cite{Hartich_2018,Hartich_Book,Schuss_2019,Grebenkov2020}.
}\\
\indent \emph{Conclusion.---}Leveraging spectral analysis and the framework of
concentration inequalities we derived general upper and lower bounds on
the probability that the empirical first-passage
time $\ntau$ inferred from $n$ independent realizations deviates from
the true mean $\mtau$ by any given amount.~Using these bounds we
constructed non-asymptotic confidence intervals that hold in the
elusive small-sample regime and thus go beyond Central-Limit- and
bootstrapping-based methods which 
fail for small $n$.~The results require
minimal input and in particular do not require any prior belief
as in the Bayesian approach that is known to be
problematic 
and likely underestimates the 
uncertainty in sub-sampling settings.~Our concentration-based results
{and bounds on extreme deviations} allow for rigorous, model-free
error control and 
reliable error estimation,
which is essential for the
analysis of experimental and simulation data.~They may further be extended to 
non-ergodic and 
irreversible dynamics.

\emph{Acknowledgments.---}Financial support from Studienstiftung des Deutschen Volkes (to R.B.)
and the German Research Foundation (DFG) through the Emmy Noether Program GO 2762/1-2 (to A.G.) is gratefully acknowledged.

\blue{\emph{Appendix~Ia:~Equilibrium initial conditions.---}We mainly consider that the initial value $x_0$ 
of the first-passage process is drawn from a ``quasi'' stationary 
equilibrium density $\tilde{p}_{\rm eq}(x)$.
However, compared to standard relaxation processes with ``true'' stationary density $p_{\rm eq}(x)$, 
an appropriate definition of  $\tilde{p}_{\rm eq}(x)$
for absorption (i.e., first-passage) processes 
is more subtle as the absorbing target $a$ has to be accounted for.
In discrete state-space we have $\tilde{p}_{\rm
  eq}(x_{k\ne a})\equiv p_{\rm
  eq}(x_k)/\sum_{k\ne a}p_{\rm
  eq}(x_k)$, i.e., 
the quasi-stationary density $\tilde{p}_{\rm eq}(x)$ is obtained
via renormalization of $p_{\rm eq}(x)$ by \emph{excluding} the target $a$.
On the contrary, in the continuous state-space setting, the
absorbing state $a$ has nominally zero measure such that trivially
$\tilde{p}_{\rm eq}(x) = p_{\rm eq}(x)$ remains unchanged.
}

\blue{\emph{Appendix~Ib:~Arbitrary initial conditions.---} 
Unless the state space $\Omega$ is both discrete and finite, an extension to arbitrary initial conditions, where $x_0$ is drawn from a general density
$p_0(x)$, requires some additional assumptions about the generalized
potential $\varphi(x)$. In particular, when $p_0(x)\ne \tilde{p}_{\rm eq}(x)$, $\varphi(x)$ must be 
sufficiently confining to ensure a
``nice'' asymptotic growth of eigenvalues $\nu_k$ of $\hat{L}$ of the
relaxation dynamics \cite{Note4}, i.e., 
$\lim_{k\to\infty}\nu_k= b k^{\beta}$ with $\beta>1/2$ and $0<b<\infty$. 
The latter condition is automatically satisfied when $\Omega$ is finite, 
since regular Sturm-Liouville problems
display Weyl asymptotics with $\beta=2$~\cite{Teschl}. 
The condition is in fact
satisfied by most physically relevant processes with discrete spectra, including the 
(Sturm-Liouville irregular)
Ornstein-Uhlenbeck or Rayleigh process~\cite{Gardiner} with
$\beta=1$.} 

\blue{Manifestations of general initial conditions $p_0(x_0)\neq \tilde{p}_{\rm eq}(x_0)$
are then fully accounted for by simply replacing
$\langle\tau^2\rangle$ with $2\sum_i w_i \mathbbm{1}_{w_i >0} /
\mu_i^2  <\infty$ in the presented results, i.e.,\ $\C\to2\sum_i w_i \mathbbm{1}_{w_i >0} (\mu_1/\mu_i)^2$.
Additional details, such as a proof of convergence of the sum $\sum_i
w_i \mathbbm{1}_{w_i >0} /\mu_i^2$,  are given in~\cite{Note5}.
}

\blue{
\emph{Appendix II:~Lower bounds on deviation probabilities.---}
There exists a ``noise
floor'' in the estimate $\ntau$ for any $n$.
Since $\mu_k\leq \mu_{k+1}$
and the weights are non-negative $w_k\ge 0$~\cite{Note5}
and normalized~\cite{Hartich_2018, Hartich_2019},
the survival probability
obeys the bound
$w_1 \e^{-\mu_1t} \leq S_a(t) \leq \e^{-\mu_1 t}$.
Using that 
$\ntau \geq \tau_n^{\rm -}$ and 
$\ntau \leq \tau_n^{\rm +}$ it follows immediately that 
$\mathbbm{P}(\mintau\geq t)\leq\mathbbm{P}(\ntau\geq t) \leq
\mathbbm{P}(\maxtau\geq t)$. 
Furthermore, we have
$\mathbbm{P}(\mintau \geq t) = S_a(t)^n$ 
as well as $\mathbbm{P}(\maxtau \leq t) = (1-S_a(t))^n$
such that we arrive at the lower bounds
\begin{align}
\mathbbm{P}\left({\ntau-\langle\tau\rangle \geq t}\right)
&\geq 
\left(w_1 \e^{-\mu_1(\mtau+t)}\right)^n\equiv\mathcal{L}_n^+(t)
\nonumber\\
\mathbbm{P}\left({\ntau-\langle\tau\rangle\leq -t}\right)
&\geq 
\left(1-\e^{-\mu_1(\mtau-t)}\right)^n\equiv\mathcal{L}_n^-(t).
\label{lower}
\end{align}
We remark that analogous results are also obtained 
for upper bounds (see \cite{Note5}) which, however, are 
weaker than those derived above 
in Eqs.~\eqref{eq:conc_n1} and~\eqref{eq:c_bound}
with the 
Cram\'er-Chernoff approach and concurrently require more information about the
dynamics.
}

\blue{
\emph{Appendix~III:~Sketch of proof of extreme-deviation bounds
and their saturation.---}In essence, 
the proofs of ``squeeze'' bounds
$\underline{\mathcal{M}}_n^{\pm}\leq \langle m^\pm_n\rangle \leq \widebar{\mathcal{M}}_n^{\pm}$
rely on bounding the average value of the minimum
$\tau^-_n\equiv\min_{i\in [1,n]}\tau_i$ or 
maximum $\tau^+_n\equiv\max_{i\in [1,n]}\tau_i$ 
out of $n$ first-passage times.  
As a first step recall that $\tau_i$'s are i.i.d., such that we may write
$\langle\tau_n^\pm\rangle=\int_0^\infty\mathbbm{P}(\tau^\pm_n >t)dt$
solely in terms of the survival probability $S_a(t)$, that is, 
\begin{align}
\langle\tau^+_n\rangle &= \int_0^\infty [1-(1-S_a(t))^n] dt
\nonumber
\\
\langle\tau^-_n\rangle &= \int_0^\infty S_a(t)^n dt.
\label{eq:sketch}
\end{align}
This enables us to fully exploit the mathematical structure 
of the spectral decomposition of $S_a(t)$, allowing us to ultimately 
arrive at Eq.~\eqref{eq:MaxMinBounds} by bounding the
respective integrands in Eq.~\eqref{eq:sketch}.}

\blue{To briefly outline the key ideas, bounds
$\underline{\mathcal{M}}_n^-$ and $\widebar{\mathcal{M}}_n^+$ 
follow by recognizing (since $\wp_a(t|\tilde{p}_{\rm eq})$ is
monotonic in $t$)
that $\langle\tau^+_n\rangle$ is maximal and $\langle\tau^-_n\rangle$
is minimal, respectively, whenever the contribution $w_1/\mu_1$ of the longest first-passage time-scale is
maximal.
For any fixed value of $\mu_1\mtau\in(0,1]$, this condition is generally met
in the presence of a spectral gap $\mu_1 / \mu_k \to 0, \forall k>1$, for which one 
may utilize in Eq.~\eqref{eq:sketch} the ansatz
\begin{align}
S_a(t) = \lim_{\varepsilon\to 0} (1-\mu_1\mtau)\e^{-t/\varepsilon} + \mu_1\mtau\e^{-\mu_1 t}.
\end{align}
The announced bounds are consequently obtained after some
straightforward calculations and become saturated (i.e.\ the
inequality becomes an equality) in systems with the
above stated spectral gap~\cite{Note5}.\\
\indent The main idea behind proving
the bound $\underline{\mathcal{M}}_n^+$ 
relies on the inequality $S_a(t)\geq w_{k_+}\e^{-\mu_{k_+} t}$,
were $k_+\equiv{\rm argmin}_k w_k >0$ denotes the smallest $k$ for which the corresponding 
weight $w_k$ is strictly positive. Saturation occurs when $w_{k_+}\to
1$.\\
\indent The remaining bound $\widebar{\mathcal{M}}_n^-$ follows
directly from upper bounding the convex function $x^n$, $\forall n\geq 1$, $x\geq 0$, by means of
Jensen's inequality and is saturated when $w_k\to 1$ for some $k$ and
for degenerate systems with identical first-passage eigenvalues
$\mu_{i}=\mu,\forall i$ for some $\mu>0$.}

\bibliographystyle{apsrev4-2.bst}
\bibliography{dev}

\let\oldaddcontentsline\addcontentsline
\renewcommand{\addcontentsline}[3]{}
\let\addcontentsline\oldaddcontentsline

\clearpage
\newpage
\onecolumngrid
\renewcommand{\thesection}{S\arabic{section}}
\renewcommand{\thefigure}{S\arabic{figure}}
\renewcommand{\theequation}{S\arabic{equation}}
\setcounter{equation}{0}
\setcounter{figure}{0}
\setcounter{page}{1}
\setcounter{section}{0}

\begin{center}\textbf{Supplementary Material for:\\Controlling Uncertainty of Empirical First-Passage Times in the
  Small-Sample Regime}\\[0.2cm]
Rick Bebon and Alja\v{z} Godec\\
\emph{Mathematical {\scriptsize bio}Physics Group, Max Planck Institute for
Multidisciplinary Sciences, Am Fa\ss berg 11, 37077
G\"ottingen}\\[0.6cm]\end{center}

\begin{center}
In this Supplementary Material (SM) we present \blue{a discussion of
  issues arising when applying Bayesian inference in the
  small-sample regime}, additional background and details of the calculations, auxiliary results, numerical methods, and mathematical proofs of the claims made in the Letter.
The sections are organized in the order as they appear in the Letter.
\end{center}
\maketitle
\tableofcontents
\newpage
\section{Issues with Bayesian inference in the
  small-sample regime}
{Here we continue and extend the discussion on issues arising in
Bayesian inference in the small-sample regime 
and conclude with difficulties that may even arise in the asymptotic large-sample limit.
Bayesian methods (see e.g.~\cite{SM_gelman1995bayesian}) do
not rely on asymptotic arguments and are therefore often (in general
erroneously \cite{SM_brazzale2007applied,SM_Lista2017})
believed to readily 
alleviate the small-sample problem. Bayesian
estimates are highly
sensitive to, dependent on, and potentially
biased by, the specification of the prior
distribution, especially in the small-sample
setting~\cite{SM_gelman1995bayesian,SM_kaplan2014bayesian,SM_mcelreath2020statistical,SM_Tavakoli2017}. Due
to the prior dependence of estimates and their uncertainties, 
Bayesian methods must be treated with care when applied to small
samples~\cite{SM_McNeish2016,SM_Smid2019}
(see~\cite{SM_Ensign2009,SM_Bacallado2009,SM_bowman2013introduction,SM_Prinz2011_2,SM_TrendelkampSchroer2015}
specifically for kinetic inference) and can perform worse than
asymptotic frequentist methods~\cite{SM_McNeish2016}.

Moreover, so-called ``credible intervals''---the Bayesian analogue to
confidence intervals---have a nominally different meaning, as
they treat the estimated parameter as a random variable. 
Bayesian posterior intervals are similarly affected by limited sampling~\cite{SM_Mostofian2019}, 
i.e.\ the constructed
uncertainty estimates and their quality 
are sensitive to the choice of prior
probability~\cite{SM_brazzale2007applied,SM_Lista2017} and may likely
underestimate the true
uncertainty and thus fail to provide trustworthy confidence 
intervals~\cite{SM_Chodera2010,SM_bowman2013introduction}.

On a more subtle level, the classical Bernstein-von-Mises
theorem establishes a rigorous (frequentist) justification
of posterior-based Bayesian credible intervals as asymptotically
correct, prior independent
confidence intervals for (finite dimensional) parametric models in the
\emph{large-sample} 
limit~\cite{SM_le2000asymptotics,SM_van2000asymptotic,SM_le2012asymptotic}. 
Analogous statements for semi-parametric and (infinite
dimensional) non-parametric models, however, 
involve more delicate conceptual and mathematical issues~\cite{SM_Diaconis1986,SM_Cox1993,SM_diaconis1998consistency,SM_freedman1999wald}
and, despite having received significant attention
\cite{SM_barron1999consistency,SM_ghosal1999posterior,SM_ghosh2003,SM_Kim2004,SM_boucheron2009bernstein,SM_bickel2012semiparametric,SM_rivoirard2012bernstein,SM_castillo2014bernstein,SM_Rousseau2016,SM_rockova2020semi,SM_Ray2021,SM_ghosal2017fundamentals}
(see also~\cite{SM_Kleijn2012} for misspecified and high
dimensional~\cite{SM_Johnstone2010} parametric models), seem to
remain---even in the asymptotic, large-sample regime---an
elusive problem.
}

\section{Spectral representation and preparatory Lemmas}
In this section we provide 
additional background on the spectral analysis of first-passage
problems and some auxiliary Lemmas. In particular, we
prove that for equilibrium initial conditions
all spectral first-passage weights $w_k(\tilde{p}_{\rm eq})$ are non-negative
and that for general initial conditions $p_0(x_0)$ the sum of positive
spectral weights is always bounded.

\subsection{Spectral representation and general results}
\label{sec:SM_spectral}
First, we recall some general results using the 
spectral representation of first-passage processes
(for more details see
e.g.~\cite{SM_Hartich2018,SM_Hartich2019}). As stated in the Letter, we consider time-homogeneous Markov
processes $x_t$ on a continuous or discrete state-space $\Omega$ with (forward) generator
$\hat{L}$ corresponding to a Markov rate-matrix or an effectively
one-dimensional Fokker-Planck
operator. 
Let the transition probability density to find $x_t$ at $x$ at time $t$ given that
it evolved from $x_0$ be $p_t(x|x_0)\equiv
\mathrm{e}^{\hat{L}t}\delta_{x_0}(x)$ where $\delta_{x_0}(x)$ denotes
the Dirac or Kronecker delta for continuous and discrete state-spaces,
respectively. We assume the process to be ergodic
$\lim_{t\to\infty}p_t(x|x_0)=p_{\rm eq}(x)$, where $p_{\rm
  eq}(x)\equiv\mathrm{e}^{-\varphi(x)}$ denotes
the equilibrium probability density and $\varphi(x)$ the corresponding
generalized potential in units of thermal energy $k_{\rm B}T$.
We assume that $\hat{L}$ 
obeys detailed balance, such that it is self-adjoint
in the left eigenspace with respect to a scalar product weighted by $\mathrm{e}^{-\varphi(x)}$
and the operator
$\mathrm{e}^{\varphi(x)/2}\hat{L}\mathrm{e}^{-\varphi(x)/2}$ is
self-adjoint with respect to a flat measure. 
 
We assume that $\hat{L}$
 is either 
(i) bounded,
(ii) $\Omega$ is finite with reflecting boundary $\partial \Omega$, 
or that (iii) $\Omega$ is infinite but $\varphi(x)$ is
sufficiently
confining (precisely, we require that $\varphi(x)$ 
satisfies the
Poincar\'e inequality,
i.e.\ $\lim_{|x|\to\infty}(|\nabla\varphi(x)|^2/2-\nabla^2\varphi(x))=\infty$).~Each
of the conditions (i)-(iii) ensures that the eigenvalue
spectrum
of $\hat{L}$ is discrete. The relaxation eigenvalue
problem (for the inner product $(\cdot|\cdot)$ defined with respect to a flat Lebesgue measure) reads $-\hat{L}\Psi^{\rm R}_k(x)=\nu_k\Psi^{\rm R}_k(x)$ with $\Psi_k^{\rm L}(x)=\Psi^{\rm R}_k(x)\mathrm{e}^{\varphi(x)}$, $\nu_0=0$
and $\nu_{k\ge 1}>0$. 

The first-passage time 
to a target $a$ for
$x_{t=0}$ drawn from a density $p_0(x_0)$ is defined as $\tau=\inf_t
[\,t\, |x_t= a, \, p_0(x_0)]$. We will use $\langle\cdot\rangle$ to
denote an average over all first-passage paths $\{x_{t'}\}_{0\le t'
  \le \tau}$, i.e.\ those that hit
$a$ only once.  The first-passage time density to
$a$, $\wp_a(t|x_0)=\langle \delta(t-\tau[\{x_{t'}\}]) \rangle$ to
reach the absorbing target at $x=a$, starting initially from $x_0$,
has the general spectral representation 
\begin{equation}
\wp_a(t|x_0) = \sum_{k\geq 1} w_k{(x_0)}\mu_k\e^{-\mu_k t},
\label{SM_FPT_density}
\end{equation}
where $\mu_k$ is the $k$-th first-passage rate and $w_k(x_0)$ its corresponding first-passage weight~\cite{SM_Hartich2018,SM_Hartich2019}.
In similar fashion the survival probability is expressed
as
\begin{equation}
S_a(t|x_0)\equiv\int_t^\infty\wp_a(t'|x_0)dt' = \sum_{k\geq 1}w_k(x_0)\e^{-\mu_k t}.
\label{SM_S_Prob}
\end{equation}
We note that in contrast to the relaxation eigenvalues $\nu_k$, the
first-passage rates $\mu_k=\mu_k(a)$ depend in the location of the absorbing
target. Moreover, for any target location $a$ the \emph{interlacing
theorem} holds~\cite{SM_Hartich2018,SM_Hartich2019} :
\begin{equation}
\nu_{k-1}\le \mu_k(a)\le \nu_k\quad \forall k,a
  \label{interlacing}
  \end{equation}
where equality occurs iff $w_k(x_0)=0$, i.e.\ for $a$ where
$\Psi_k^{\rm R}(a)=0$. 

Laplace transforming
the spectral expansion of the first-passage time density~\eqref{SM_FPT_density}---according to
$\tilde{f}(s)\equiv\int\e^{-st} f(t)\, dt$ with $f$ being a generic
function locally integrable on $t\in[0,\infty)$---yields
\begin{equation}
\tilde{\wp}_a(s)=\sum_{k\geq 1}\frac{w_k(x_0)\mu_k}{s+\mu_k}.
\end{equation}
The first-passage weights are then obtained by 
using the residue theorem to invert the Laplace 
transformed renewal theorem~\cite{SM_Siegert1951,SM_Hartich2018,SM_Hartich2019}
\begin{equation}
w_k(x_0)
=\frac{\tilde{p}(a,-\mu_k|x_0)}{\mu_k\dot{\tilde{p}}(a,-\mu_k|a)}
=\frac{\sum_{l\ge 0}(1-\nu_l/\mu_k)^{-1}\Psi_l^{\rm R}(a)\Psi_l^{\rm L}(x_0)}{\sum_{l\ge 0}(1-\nu_l/\mu_k)^{-2}\Psi_l^{\rm R}(a)\Psi_l^{\rm L}(a)}<\infty,
\label{weights}
\end{equation}
where $\dot{\tilde{p}}(a,s|a)=\partial_s \tilde{p}(a,s|a)$ is taken at $s=-\mu_k$
and $\{\nu_l,\Psi_l^{\rm R}, \Psi_l^{\rm L}\}$ are the corresponding 
\emph{relaxation eigenmodes} \cite{SM_Hartich2018,SM_Hartich2019}. 
The weights satisfy $\sum_{k\geq 1} w_k(x_0)=1$ 
and the first non-zero weight is strictly positive $w_1(x_0)>0$.
Moreover, the relaxation eigenvalues $\nu_0=0$ and all 
$\nu_{k>0}\ge 0$ are real as a result of detailed balance.

\subsection{Lemma 1: All weights are non-negative for equilibrium initial conditions}
In the Letter we focus on \emph{equilibrium} initial conditions, 
that is we assume that $x_0$ is drawn from the invariant measure,
$p_{\rm eq}(x_0)$, which in the particular case of diffusion processes
is assumed to have a reflecting boundary at $a$ (i.e.~we focus on the
one-sided first-passage process). 
We further introduce the non-negative modified spectral weights 
$\bar{w}_k(x_0)\equiv w_k(x_0)\theta(\sgn[w_k(x_0)])$
and now prove that for a normalized
equilibrium probability density of initial conditions $p_0(x_0)$ that excludes the
target---i.e.~$\tilde{p}_{\rm eq}(x_0)\equiv p_{\rm eq}(x_0)[1-\delta_{a}(x_0)]/
(1|p_{\rm eq}(x_0)[1-\delta_{a}(x_0)])$ where $\delta_{a}(x_0)$ is the
Dirac measure (note that $(1|\tilde{p}_{\rm eq})=1$)---all
weights $w_k$ are rendered non-negative.
We thus have $\bar{w}_k(\tilde{p}_{\rm eq})= w_k(\tilde{p}_{\rm eq})\geq 0,\forall k$. 

Namely, because $\Psi_l^{\rm L}(a)=\e^{\varphi(a)}\Psi_l^{\rm R}(a)$ 
we have $\Psi_l^{\rm R}(a)\Psi_l^{\rm L}(a)\ge 0, \forall l$, and from
bi-orthogonality $(\Psi_l^{\rm L}|p_{\rm eq})=\delta_{l,0}$ it
follows that [with $\tilde{p}_{\rm eq}(a) = p_{\rm eq}(a)/(1-p_{\rm eq}(a))$]
\begin{equation}
  \tilde{w}_k\equiv (w_k|\tilde{p}_{\rm eq})=\tilde{p}_{\rm eq}(a)\frac{1-\sum_{l\ge
    0}(1-\nu_l/\mu_k)^{-1}\Psi_l^{\rm R}(a)\Psi_l^{\rm L}(a)}{\sum_{l\ge
    0}(1-\nu_l/\mu_k)^{-2}\Psi_l^{\rm R}(a)\Psi_l^{\rm L}(a)}=\frac{\tilde{p}_{\rm eq}(a)}{\sum_{l\ge
    0}(1-\nu_l/\mu_k)^{-2}\Psi_l^{\rm R}(a)\Psi_l^{\rm L}(a)}\ge 0,
  \label{eweights}
\end{equation}
because by definition $\mu_k>0,\forall k\ge 1$ denotes the zeros of
$\tilde{p}(a,s|a)$, i.e. $\tilde{p}(a,-\mu_k|a)=\sum_{l\ge
    0}(\nu_l-\mu_k)^{-1}\Psi_l^{\rm R}(a)\Psi_l^{\rm L}(a)=\mu_k^{-1}\sum_{l\ge
    0}(1-\nu_l/\mu_k)^{-1}\Psi_l^{\rm R}(a)\Psi_l^{\rm L}(a)=0$
    which completes the proof of the Lemma. 

\subsection{Lemma 2: Sum of positive weights is bounded from above}

For the sake of completeness we present supplementary results for
general initial conditions $p_0(x_0)$. Recall from the Letter that we
require some additional conditions on $\varphi(x)$ or $\Omega$ in this more
general setting.

In particular, we assume that $\varphi(x)$ is sufficiently confining to assure a
``nice'' asymptotic growth of eigenvalues, $\lim_{k\to\infty}\nu_k= b
k^{\beta}$ with $\beta>1/2$ and $0<b<\infty$. The latter condition is
automatically satisfied when $\Omega$ is finite, since regular
Sturm-Liouville problems display Weyl asymptotics with
$\beta=2$~\cite{SM_Teschl}. The condition is in fact satisfied by most
physically relevant processes with discrete spectra, including the
(Sturm-Liouville irregular) Ornstein-Uhlenbeck or 
Rayleigh process \cite{SM_Gardiner} with $\beta=1$. This implies, by
the interlacing theorem \eqref{interlacing} that $b
(k-1)^{\beta}\le \mu_k\le b k^{\beta}$ and therefore there exists a real
constant $C\in(0,\infty)$ such that
$\lim_{k\to\infty}\mu_k$ diverges as $Ck^\beta$.

Recall further that the $m$-th moment of $\tau$ is given 
by $\langle\tau^m\rangle = m! \sum_{k\geq 1} w_k(p_0)/\mu_k^m$.
By construction we obtain
$2\sum_{k\ge 1}\bar{w}_k(p_0)/\mu_k^2\equiv \langle
\bar{\tau}_{p_0}^2\rangle \ge 2\sum_{k\ge
  1}w_k(p_0)/\mu_k^2\equiv \langle \tau^2_{p_0}\rangle$, where
equality holds when $p_0=\tilde{p}_{\rm eq}$ (since in this case all $w_k\geq 0$, i.e., $\bar{w}_k(\tilde{p}_{\rm eq}) = w_k(\tilde{p}_{\rm eq})$ 
as discussed before).

Moreover, because we only
consider Markov jump processes on finite state-spaces as well as processes for which $\lim_{k\to\infty}\mu_k=C
k^{\alpha}$ with $0<C<\infty$ and $\alpha>1/2$ (this includes confined Markov
jump processes on infinite state-spaces and all regular
Sturm-Liouville problems) convergence is ensured, i.e.~$2\sum_{k\ge
  1}\bwk/\mu_k^{2+n}<\infty,\forall n\ge 0$.

To prove this consider $w_{\rm max}\equiv
\max_{k\ge k_*} \bwk$ such that $w_{\rm max}/\mu_k^{2+n}\ge
w_k(p_0)/\mu_k^{2+n},\forall k$. Let the smallest $k$ for which the asymptotic
scaling holds be $k_*$ then we may split the summation as $\sum_{k\ge
  1}=\sum_{k=1}^{k_*-1}+\sum_{k\ge k_*}$ such that
\begin{equation*}
\sum_{k\ge 1}\frac{\bwk}{\mu_k^{2+n}}\le\sum_{k=1}^{k_*-1}\frac{\bwk}{\mu_k^{2+n}}+\sum_{k\ge k_*}\frac{w_{\rm max}}{\mu_k^{2+n}}.
\end{equation*}
Because the first term is nominally finite we only need to prove
convergence of the second sum, which we do by means of the integral test. We
define a function $f(k)\equiv w_{\rm max}/\mu_k^{2+n}$ that is
monotonically decaying in $k$. This implies $f(x)\le f(k),\forall
x\in[k,\infty)$ and $f(x)\ge f(k),\forall
x\in[k_*,k]$. We then have for every integer $k\ge k_*$ that
$\int_{k}^{k+1}f(x)dx\le\int_{k}^{k+1}f(k)dx=f(k)$ and conversely,  for every integer $k\ge k_*+1$ that
$\int_{k-1}^{k}f(x)dx\ge\int_{k-1}^{k}f(k)dx=f(k)$. We now sum over
all $k\ge k_*$ to obtain, using $\mu_k=C
k^{\alpha}\forall k\ge k_*$
\begin{eqnarray*}
\int_{k_*}^{\infty}\frac{w_{\rm max}}{(Cx^{\alpha})^{(2+n)}}dx\le
\sum_k\frac{w_{\rm max}}{\mu_k^{2+n}}&\le& \frac{w_{\rm
    max}}{(Ck_*^{\alpha})^{2+n}}+\int_{k_*}^{\infty}\frac{w_{\rm
    max}}{(Cx^{\alpha})^{2+n}}dx\to\\
\frac{w_{\rm max}C^{-(2+n)}k_*^{1-\alpha(2+n)}}{\alpha(2+n)-1}\le
\sum_k\frac{w_{\rm max}}{\mu_k^{2+n}}&\le&w_{\rm
    max}(Ck_*^{\alpha})^{-(2+n)}+\frac{w_{\rm max}C^{-(2+n)}k_*^{1-\alpha(2+n)}}{\alpha(2+n)-1}<\infty
\end{eqnarray*}  
where the last integral converges because $1-\alpha(2+n)<0,\forall
n\ge 0$, which in turn proves convergence of $\sum_{k\ge
  1}\bwk/\mu_k^2$.

\section{Extreme value bounds and comparison with Cram\'er-Chernoff bounds}
In the Letter we derive lower bounds $\mathcal{L}^\pm_n(t)$ on 
the deviation probability $\mathbbm{P}(\ntau-\mtau\ge t)$ 
and $\mathbbm{P}(\mtau-\ntau\ge t)$ 
by utilizing \emph{extremal events}, i.e.,
we consider the \emph{maximal} and \emph{minimal} first-passage time 
in a sample of $n\ge 1$ i.i.d.~realizations.
In this section we derive analogous \emph{upper bounds}
building on the same ideas.

\subsection{Extreme value bounds for the sample mean $\ntau$}

Recall that for the reversible Markov dynamics considered the equilibrium 
survival probability
$S_a(t|\tilde{p}_{\rm eq})\equiv S_a(t)$
in its spectral representation~\eqref{SM_S_Prob} 
obeys
\begin{equation}
w_1 \e^{-\mu_1 t}\leq S_a(t)\leq\e^{-\mu_1 t}.
\label{SM_SBound}
\end{equation}
For the upper bound we use $\mu_k\leq\mu_{k+1}$ and that $\sum_{k>0} w_k=1$ are normalized,
whereas the lower bound follows since $w_k\geq 0$, $\forall k$,
as we consider
equilibrium initial conditions throughout.
Moreover, from extreme value theory it follows
\begin{align}
\mathbbm{P}(\mintau\geq t) = S_a(t)^n
\quad&\Leftrightarrow\quad
\mathbbm{P}(\mintau\leq t) = 1-S_a(t)^n,
\nonumber
\\
\mathbbm{P}(\maxtau\leq t)=\left(1-S_a(t)\right)^n
\quad&\Leftrightarrow\quad
\mathbbm{P}(\maxtau\geq t)=1-\left(1-S_a(t)\right)^n,
\label{SM_EV}
\end{align}
where we introduce $\maxtau\equiv\max_{i\in[1,n]}\tau_i$ 
and $\mintau\equiv\min_{i\in[1,n]}\tau_i$, respectively. 
Clearly, since $\mintau\leq\ntau\leq\maxtau$ we may write
$\mathbbm{P}(\mintau\geq t)\leq\mathbbm{P}(\ntau\geq t)\leq\mathbbm{P}(\maxtau\geq t)$
and analogously
$\mathbbm{P}(\mintau\leq t)\geq\mathbbm{P}(\ntau\leq t)\geq\mathbbm{P}(\maxtau\leq t)$.
Using
Eq.~\eqref{SM_EV} in combination with Eq.~\eqref{SM_SBound} we directly 
arrive at the \emph{lower bounds} $\mathcal{L}^\pm_n(t)$ (see Eq.~(4) in the Letter)
\begin{align}
\mathbbm{P}(\ntau\geq\mtau +t)
&\geq\mathbbm{P}(\mintau\geq\mtau +t)
= S_a(t+\mtau)^n
\geq \left(w_1\e^{-\mu_1(\mtau +t)}\right)^n
\\
\mathbbm{P}(\ntau\leq\mtau-t)
&\geq \mathbbm{P}(\maxtau\leq \mtau -t)
=\left(1-S_a(\mtau -t)\right)^n
\geq
\left(1-\e^{-\mu_1(\mtau -t)}\right)^n.
\label{SM_EV_Lower}
\end{align}
Introduced considerations are, however, not restricted 
to only lower bounds such that we can further leverage bounds on the equilibrium survival probability~\eqref{SM_SBound} 
to analogously obtain 
corresponding \emph{upper bounds} as 
\begin{align}
\mathbbm{P}(\ntau\geq\mtau +t)
&\leq\mathbbm{P}(\maxtau\geq \mtau +t)
=1-(1-S_a(\mtau +t))^n
\leq 1-\left(1-\e^{-\mu_1(\mtau + t)}\right)^n,
\nonumber
\\
\mathbbm{P}(\ntau\leq\mtau-t)
&\leq\mathbbm{P}(\mintau\leq\mtau - t)
=1-S_a(\mtau -t)^n
\leq 1-\left(w_1\e^{-\mu_1(\mtau -t)}\right)^n.
\label{SM_EV_Upper}
\end{align}
As we will illustrate next, the upper bounds for the sample mean~\eqref{SM_EV_Upper}
are much weaker than those derived with the Cramér-Chernoff approach (Eq.~(7) in the Letter)
and require more information about the dynamics.

\subsection{Comparison of Cram\'er-Chernoff vs Extreme value Bounds}

\begin{figure}[htbp]
\centering
\includegraphics{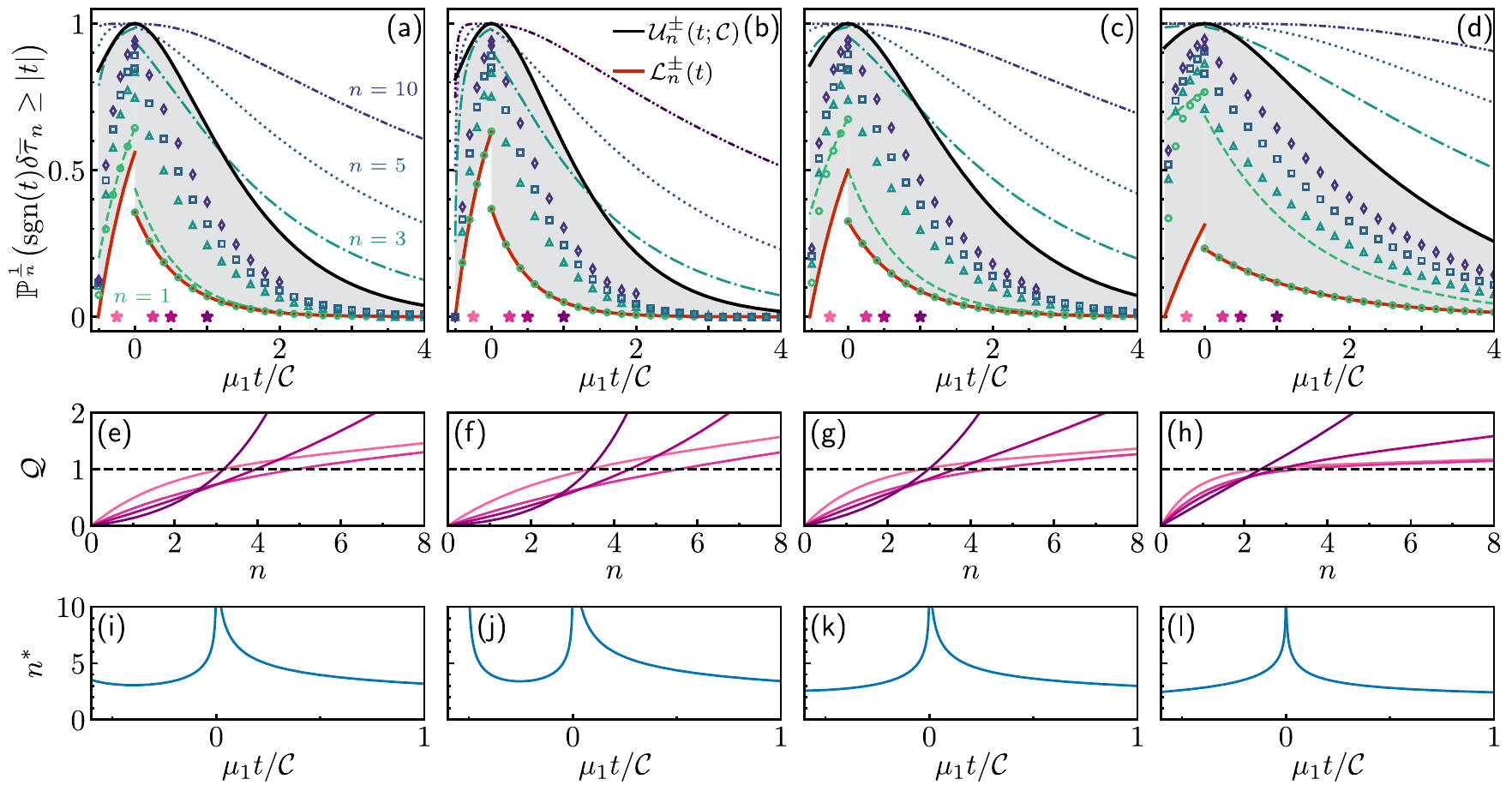}
\caption{Comparison between Cram\'er-Chernoff-type upper bounds $\mathcal{U}^\pm_n(t;\C)$ and extreme value upper bounds 
for a spatially confined Brownian search process in dimensions (a,e,i) $d=1$
and (b,f,j) $d=3$, and discrete-state Markov jump processes
for (c,d,k) the inferred model of calmodulin and (d,h,l) a 8-state toy protein.
(a-d) Scaled probabilities $\mathbbm{P}^{1/n}({\rm sgn}(t)\delta\ntau\geq |t|)$ that the sample mean $\ntau$ inferred from $n\geq 1$ realizations
deviations from $\mtau$ by more than $t$ in either direction. 
Right tail areas are shown for $t>0$ and left for $t<0$, respectively.
Cram\'er-Chernoff upper bounds $\mathcal{U}^\pm_n(t;\C)$ are displayed as black and extreme value upper bounds as dashed lines, respectively.
Corresponding lower bounds $\mathcal{L}^\pm_n(t)$ are depicted as red lines and symbols denote scaled empirical deviation probabilities obtained
from the statistics of $\ntau$ for different $n$.
(e-h) Quality factor $\mathcal{Q}$ as a function of $n$ for different fixed relative deviations $\mu_1 t$ (see star symbols (a-d)).
(i-l) Sample size $n^*$ (blue) for which both upper bounds are equal, i.e., $\mathcal{Q}=1$,
as a function of re-scaled deviations.
}
\label{fig:SM_comparison}
\end{figure}

In this section we directly compare the 
concentration-based upper bounds $\mathcal{U}^\pm_n(t)$ (see Eq.~(7) in the Letter)
that are obtained with the Cramér-Chernoff approach,
with the upper bounds~\eqref{SM_EV_Upper} which are based on extreme value considerations
in analogy to the lower bounds $\mathcal{L}^\pm_n(t)$.
Similar to Fig.~2e-h of the Letter 
we now exemplify and compare both upper bounds  
in Fig.~\ref{fig:SM_comparison} 
for the model systems shown in Fig.~1b-d.

In Fig.~\ref{fig:SM_comparison}a-d we equivalently express 
re-scaled deviation probabilities
$\mathbbm{P}^{1/n}(\text{sgn}(t)\delta\ntau\geq |t|)$ 
in a single panel, i.e., for
the left tail we formally let $t\to -t$ such 
that $t$ as shown now has support in $[-\mtau, \infty)$
and $\text{sgn}(x)=\pm 1$ for $\pm x >0$ and $\text{sgn(0)}=0$
denotes the signum function.
Empirical deviation probabilities (symbols) as a function of $t$ are computed from statistics obtained by sampling 
$\ntau$ for different fixed $n$ values.
Extreme value lower bounds $\mathcal{L}^\pm_n(t)$~\eqref{SM_EV_Lower} for both tails are depicted in red.
Here we now focus on comparing the \emph{upper bounds}.
Concentration inequalities $\mathcal{U}_n^\pm(t;\C)$ (Eq.~(7))
are again depicted as black lines
whereas 
the corresponding extreme value upper bounds are represented as dashed/dotted lines where the respective 
coloring indicates the number of realizations $n$.
Note, that the concentration bounds (and the lower bounds) collapse onto a 
single master curve due to the employed scaling $\mathbbm{P}^{1/n}$,
whereas the extreme value upper bounds do \emph{not}
due to their different functional form (compare Eq.~\eqref{SM_EV_Upper}).
Evidently, while for $n=1$  the extreme value bounds remains close to the actual deviation probability,
already for $n=3$ they become considerably less tight and overshoot heavily 
for all considered models.
Moreover, extreme value upper bounds become increasingly weak (even trivial at times) as $n$ increases,
therefore highlighting that Cram\'er-Chernoff-type bounds
are vastly more suitable.

Motivated by the discussion above we next want to gain more quantitative insights for which 
sample sizes $n$ the Cramér-Chernoff approach becomes more favorable.
For this purpose we introduce a \emph{quality factor} 
$\mathcal{Q}\in[0,\infty)$
that is informally defined as 
\begin{equation}
\mathcal{Q}\equiv \frac{\text{Extreme value upper bound}}{\text{Cram\'er-Chernoff-type upper bound}}.
\end{equation}
A value $\mathcal{Q}>1$ therefore indicates that the Cram\'er-Chernoff 
bound is tighter and $\mathcal{Q}<1$ suggests
that the extreme value bound should be favored, respectively.
In Fig.~\ref{fig:SM_comparison}e-h we illustrate the quality factor $\mathcal{Q}$ 
as a function of sample size $n$ for different fixed dimensionless deviation values 
$\mu_1 t$ (star symbols in Fig.~\ref{fig:SM_comparison}a-d).
Remarkably for all model systems considered---which span a large range of possible $\mathcal{C}$ 
values---the Cram\'er-Chernoff approach is already superior 
even in the small-sample regime  $n\lesssim 4$.
Moreover, we can further study the particular $n^*$, for which one would reach
$\mathcal{Q}=1$, as a function of some desired deviation $\mu_1 t$
relative to the longest time scale $1/\mu_1$.
Note, that again for the left tail we let $t\to -t$ (see discussion above).
As depicted in Fig.~\ref{fig:SM_comparison}i-l for our model systems,
$n^*$ (blue) generally is found to be well below $n=8$, i.e., even for most small sample sizes
the derived Cramér-Chernoff-type bounds can be considered to be the better choice, especially when considering large $\mu_1 t$ (i.e.~large deviations).

Lastly, one could ask the question why the extreme value upper bound is so ``weak'' when $n$ increases even just slightly.
To answer this question we recall 
that---since we are interested in deviations of the sample mean $\ntau$ around $\mtau$---we bound the sample mean
with the minimal and maximal first-passage time according to 
$\mintau\leq\ntau\leq\maxtau$ which is further used,
in combination with bounds on the survival probability~\eqref{SM_SBound},
to derive corresponding upper bounds~\eqref{SM_EV_Upper}.
Clearly, as $n$ increases 
we expect this bound to become increasingly loose as by larger sample sizes
we increase the chances of sampling rare first-passage times, i.e., maximal and minimal first-passage time
that strongly deviate from the (sample) mean---this also explain why bounds~\eqref{SM_EV_Lower}
and~\eqref{SM_EV_Upper} are only particularly tight for $n=1$ as here $\mintau =\overline{\tau}_1 =\maxtau$.
In contrast, the Cramér-Chernoff method requires a much more delicate mathematical analysis 
involving bounds of the moment generating function.
The Cram\'er-Chernoff-type 
bound has the additional advantage that it can be further used to \emph{universally} bound deviation probabilities where
no specific information about the underlying system is required (see Eq.~(11) in the Letter).
Moreover, even the version of Cram\'er-Chernoff bounds $\mathcal{U}^\pm_n(t;\C)$ 
that require input of one system-dependent constant $\C$ still require less information about the dynamics since
extreme value upper bounds~\eqref{SM_EV_Upper} partly also require knowledge about the first-passage weight $w_1$ and $\mtau$ itself.
\newpage
\section{Complete proof of concentration inequalities and their asymptotics}
In this section we provide various additional details on the upper bounds $\mathcal{U}^\pm_n(t;\C)$ 
(Eq.~(7) of the Letter).
In particular, we prove
the required bounds on the cumulant generating function,
compute their corresponding Cram\'er transform, and give further information about the large-sample limit $n\to\infty$, as well 
as the model-free version of the bounds.
\blue{
Note that while the following proofs build on the mathematical structure 
encoded in the spectral decomposition, 
the derived bounds do not 
require knowledge of the full first-passage time spectrum $\{w_i, \mu_i \}$.
}

\subsection{Theorem 1: Cramér-Chernoff bound for the right tail $\tau \geq \mtau$}
We begin with the right tail, i.e.~upwards deviations such that $\tau\geq\mtau$,
and start by
proving a bound for the moment generating function of 
the deviation of the first-passage time $\tau$
from the mean $\langle\tau\rangle$.
Using the spectral representation~\eqref{SM_FPT_density}
and the inequality $x\leq\e^{x-1}$, $\forall x\in\mathbb{R}$,
we find 
\begin{equation}
\langle
\e^{\lambda(\tau-\mtau)}\rangle=\e^{-\lambda\mtau}\sum_{k>0}\frac{w_k}{1-\lambda/\mu_k}\le \exp\left(-\lambda\mtau+\sum_{k>0}\frac{w_k}{1-\lambda/\mu_k}-1\right)
\label{cf}
\end{equation}
for all $\lambda<\mu_k$.
Moreover, for $|\lambda|<\mu_1$ we may further expand the sum
$\sum_{k>0}\frac{w_k}{1-\lambda/\mu_k}=\sum_{m\ge0}\lambda^m\sum_{k>0}w_k/\mu_k^m$ using the geometric series.
Recall that the moments are given by
$\langle \tau^m\rangle=m!\sum_{k>0}w_k/\mu_k^m$, such that we obtain
\begin{equation}
\langle\e^{\lambda(\tau-\mtau)}\rangle
\le\exp\left(\sum_{m\ge 2}\lambda^m\sum_{k>0}\frac{w_k}{\mu_k^m}\right)
=\exp\left(\lambda^2\frac{\langle\tau^2\rangle}{2}+\sum_{m>2}\lambda^m\sum_{k>0}\frac{w_k}{\mu_k^m}\right).
\end{equation}  
Since $\mu_1\leq\mu_{k>1}$ and all first-passage weights $w_k$
are positive (due to equilibrium initial conditions) we find
\begin{align}
\langle\e^{\lambda(\tau-\mtau)}\rangle
&\le
\exp\left(\lambda^2\frac{\langle\tau^2\rangle}{2}+\sum_{m>2}\lambda^m\sum_{k>0}\frac{{w}_k}{\mu_k^m}\right)\nonumber\\
&\le
\exp\left(\lambda^2\frac{\langle\tau^2\rangle}{2}+\sum_{m>2}\frac{\lambda^m}{\mu_1^{m-2}}\sum_{k>0}\frac{{w_k}}{\mu_k^2}\right)\nonumber\\
&\le\exp\left(\lambda^2\frac{\langle\tau^2\rangle}{2}\left[1+\frac{\lambda}{\mu_1-\lambda}\right]\right)\nonumber
=\exp\left(\lambda^2\frac{\langle\tau^2\rangle/2}{(1-\lambda/\mu_1)}\right).
\label{SM_MGF_R}
\end{align}
Introducing $\psi_{\delta\tau}(\lambda)\equiv\ln\langle\e^{\lambda\delta \tau}\rangle$, with $\delta\tau= \tau-\mtau$ for 
the right tail,
we immediately identify the upper bound
\begin{equation}
\psi_{\delta\tau}(\lambda)\leq \frac{\lambda^2}{2}\frac{\langle\tau^2\rangle}{1-\lambda/\mu_1}
= \frac{\tlambda^2}{2}\frac{\C}{1-\tlambda}
\equiv\phi_{\delta\tau}(\tlambda; \C)
\qquad \tau\geq\mtau,
\label{SM_R_CGF_Bound}
\end{equation}
which concludes the derivation of the upper expression in Eq.~(6) of the Letter.
Note that we further have introduced the dimensionless quantities $\ttilde\equiv\mu_1 t$,
$\C=\mu_1^2\langle\tau^2\rangle$, and $\tlambda = \lambda/\mu_1$
in the last step. In the case of general initial conditions
$p_0(x_0)\ne \tilde{p}_{\rm eq}(x_0)$ we must simply replace
$\C\to 2 \sum_i w_i \mathbbm{1}_{w_i>0}(\mu_1/\mu_i)^2$
(see Lemma 2).

Next, we find the optimizing value of $\tlambda$, i.e., we compute the Cram\'er transform of Eq.~\eqref{SM_R_CGF_Bound}
defined as
\begin{equation}
\phi_{\delta\tau}^*(\ttilde;\C)
\equiv
\sup_{\tlambda\in[0,1)}[\tlambda \ttilde -\phi_{\delta\tau}(\tlambda;\C)]
=
\sup_{\tlambda\in[0,1)}\left[\tlambda \ttilde -\frac{\tlambda^2}{2}\frac{\C}{1-\tlambda}\right].
\label{SM_CramerT_R}
\end{equation}
$\phi_{\delta\tau}(\tlambda;\C)$ is differentiable, non-negative, convex, and increasing
on $\tilde{\lambda}\in[0, 1)$, which implies that Eq.~\eqref{SM_CramerT_R}
can be obtained by differentiation of $\tlambda \ttilde -\phi_{\delta\tau}(\tlambda;\C)$ 
with respect to $\tlambda$, hence
$\phi_{\delta\tau}^*(\ttilde;\C) = \tlambda^\dagger\ttilde - \phi_{\delta\tau}(\tlambda^\dagger;\C)$
where the optimum $\tlambda^\dagger$ solves $\phi'_{\delta\tau}(\tlambda^\dagger;\C) = t$.
Accordingly, we find the supremum to be attained at
$\tlambda^\dagger(\ttilde)=1-1/\sqrt{1+2\ttilde/\C}$.
For convenience we further introduce the auxiliary function 
$h_+(u)\equiv 1+u-\sqrt{1+2u}$ such that 
we finally arrive at 
\begin{equation}
\phi^*_{\delta\tau}(\ttilde;\C) = \C h_+(\ttilde/\C) = \C h_+(\mu_1 t/\C), \quad 0\leq t\leq\infty
\label{SM_CT_R}
\end{equation}
By using Chernoff's inequality we subsequently obtain the upper bound 
$\mathbbm{P}(\delta\ntau\geq t)\leq \e^{-n\phi^*_{\delta \tau}(t;\C)}\equiv\mathcal{U}^+_n(t;\C)$
for $0\leq t\leq\infty$
which completes the proof of Theorem 1 and thus the first announced inequality (7) in the Letter.

\begin{figure}[htbp]
\centering
\includegraphics[scale=1]{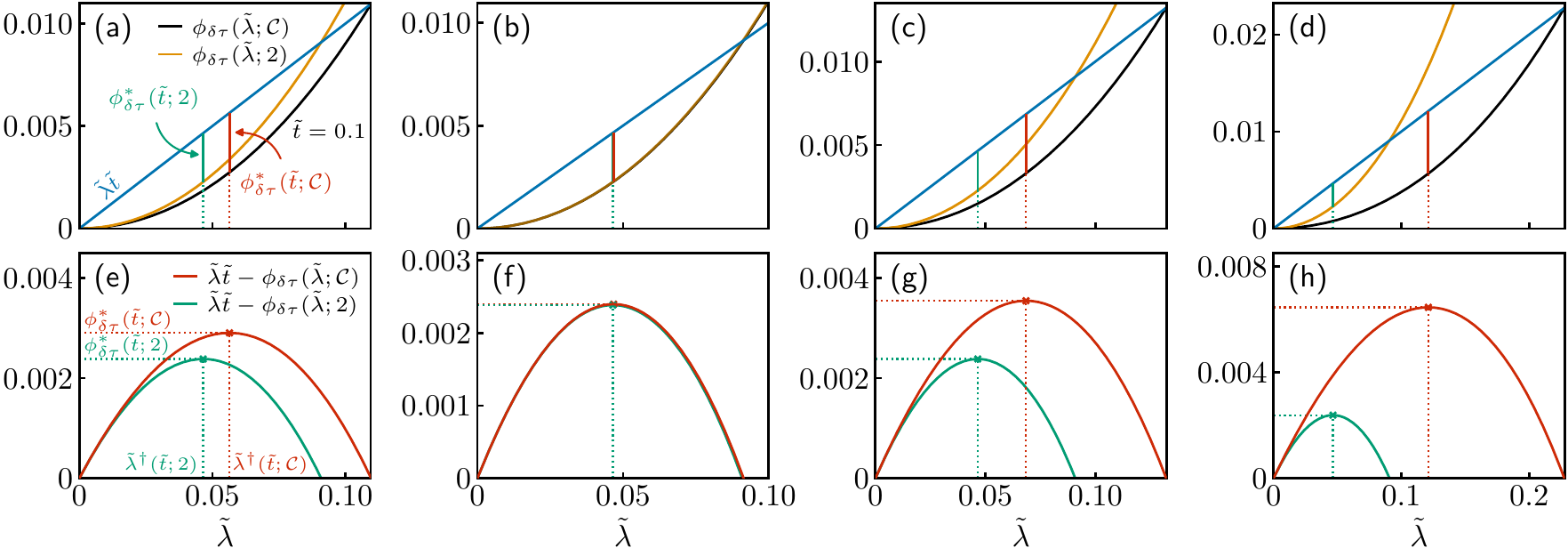}
\caption{Illustration of the Cram\'er-Chernoff bounding method for the right tail with 
$\ttilde=0.1$ and parameters for spatially confined Brownian search process in dimensions $d=1$ (a,e)
or $d=3$ (b,f), and discrete-state Markov jump processes
for the model of calmodulin (c,d) and a 8-state toy protein (d,h).
Top row depicts bounds of the cumulant generating function $\phi_{\delta\tau}(\tlambda;\C)$ (black)
and $\phi_{\delta\tau}(\tlambda;2)$ (yellow)
as a function of $\tlambda$, respectively.
Bottom row shows the differences $\tlambda\ttilde-\phi_{\delta\tau}(\tlambda;\C)$ (red)
and $\tlambda\ttilde-\phi_{\delta\tau}(\tlambda;2)$ (green) as a function of $\tlambda$, respectively (see also top row with $\tlambda\ttilde$ in blue).
The corresponding suprema are obtained at $\tlambda^\dagger(\ttilde;\C)$ and $\tlambda^\dagger(\ttilde;2)$ (dotted lines)
and define the Cram\'er transforms $\phi_{\delta\tau}^*(\ttilde;\C)$ and 
$\phi_{\delta\tau}^*(\ttilde;2)$ (compare top row).
For all considered models we demonstrate $\phi_{\delta\tau}(\tlambda;\C)\leq \phi_{\delta\tau}(\tlambda;2)$ and $\phi_{\delta\tau}^*(\ttilde;\C)\geq\phi_{\delta\tau}^*(\ttilde;2)$
as derived in the maintext.
Note for the panels (b,f) we have 
$\phi_{\delta\tau}(\tlambda;\C)\lessapprox\phi_{\delta\tau}(\tlambda;2)$
and $\phi_{\delta\tau}^*(\ttilde;\C)\gtrapprox\phi_{\delta\tau}^*(\ttilde;2)$
since $\C= 1.99\approx 2$.
}
\label{fig:SM_CT_R}
\end{figure}

\subsection{Theorem 2: Cram\'er-Chernoff bound for the left tail $\tau < \mtau $}
Next we turn to the left tail, $\tau<\mtau$,
where the corresponding moment generating function analogously reads
\begin{equation}
\langle
\e^{\lambda(\mtau-\tau)}\rangle 
= \e^{\lambda\mtau}\sum_{k>0}\frac{w_k}{1+\lambda/\mu_k}\nonumber
\le
\exp\left(\lambda\mtau+\sum_{k>0}\frac{{w}_k}{1+\lambda/\mu_k}-1\right)
\end{equation}
for $\lambda <\mu_k$.
Using equivalent arguments as for the right tail above we may further write
\begin{align}
\langle\e^{\lambda(\mtau-\tau)}\rangle 
&\leq\exp\left(\sum_{m\ge  2}(-\lambda)^m\sum_{k>0}\frac{w_k}{\mu_k^m}\right)\nonumber
\\
&=\exp\left(\lambda^2\frac{\langle\tau^2\rangle}{2}+\sum_{m>2}(-\lambda)^m\sum_{k>0}\frac{{w}_k}{\mu_k^m}\right)\nonumber\\
&\le
\exp\left(\lambda^2\frac{\langle\tau^2\rangle}{2}+\sum_{m>0}\frac{\lambda^{2m}}{\mu_1^{2m-2}}\sum_{k>0}\frac{{w}_k}{\mu_k^2}\right)
=\exp\left(\lambda^2\frac{\langle\tau^2\rangle/2}{1-(\lambda/\mu_1)^2}\right).
\label{SM_MGF_L}
\end{align}
Recall the definition 
of the cumulant generating function, $\psi_{\delta\tau}(\lambda)\equiv\ln\langle\e^{\lambda\delta \tau}\rangle$, 
such that Eq.~\eqref{SM_MGF_L} directly yields 
\begin{equation}
\psi_{\delta\tau}(\lambda) \leq \frac{\lambda^2}{2}\frac{\langle\tau^2\rangle}{1-(\lambda/\mu_1)^2}
= \frac{\tlambda^2}{2}\frac{\C}{1-\tlambda^2}
\equiv \phi_{\delta\tau}(\tlambda;\C)
\label{SM_CGF_L}
\end{equation}
which completes the derivation of the lower expression in Eq.~(6) of the Letter.
Note that for the left tail we have  $\delta\tau = \mtau - \tau$
and we again let 
$\ttilde\equiv\mu_1 t$,
$\tlambda \equiv \lambda/\mu_1$, and
$\C\equiv\mu_1^2\langle\tau^2\rangle$. In the case of general initial conditions
$p_0(x_0)\ne \tilde{p}_{\rm eq}(x_0)$ we must simply again replace
$\C\to 2 \sum_i w_i \mathbbm{1}_{w_i>0}(\mu_1/\mu_i)^2$ (see Lemma 2).

Analogous to the right tail we next compute the Cram\'er transform of Eq.~\eqref{SM_CGF_L}, i.e., 
\begin{equation}
\phi_{\delta\tau}^*(\ttilde;\C)
\equiv
\sup_{\tlambda\in[0,1)}[\tlambda \ttilde -\phi_{\delta\tau}(\tlambda;\C)]
=
\sup_{\tlambda\in[0,1)}\left[\tlambda \ttilde -\frac{\tlambda^2}{2}\frac{\C}{1-\tlambda^2}\right],
\end{equation}
where we find the optimal value $\tlambda^\dagger(\ttilde;\C)$ 
to be determined by the transcendental quartic,
$\tlambda^{\dagger}(\ttilde):(1-\tlambda^2)^2-\Ctt\tlambda=0$
with $\Ctt\equiv\C/\ttilde$, which we solve according to
the method of Descartes. 
First, we re-arrange the quartic as
$\tlambda^4-2\tlambda^2-\Ctt\tlambda+1=0$ and make the factorization ansatz
\begin{align}
  &(\tlambda^2-\y\tlambda^2+\w)(\tlambda^2+\y\tlambda^2+\z)=0\nonumber\\
  &\w+\z-\y^2=-2\nonumber\\
  &\y(\w-\z)=-\Ctt\nonumber\\
  &\z\w=1.
  \label{ansatz}
\end{align} 
The system of equations~\eqref{ansatz} is solved by
\begin{align}
  \w(\y)&=(\y^2-2-\Ctt/\y)/2\nonumber,\\
  \quad \z(\y)&=(\y^2-2+\Ctt/\y)/2,
\label{cfs}  
\end{align}
where $\y^2\equiv \Y$ is the solution of the cubic $\Y^3-4\Y^2-\Ctt^2=0$. 
Moreover, since the discriminant $\mathrm{D}$ is strictly negative,
i.e.~$\mathrm{D}=-2^8\Ctt^2-3^3\Ctt^4<0$, 
the qubic has only one real root. 
The corresponding depressed qubic reads
$\ttilde^3-2^4/3\ttilde-(2^7/3^3+\Ctt^2)=0$ with $\ttilde\Y-4/3$. 
Let $p=-2^4/3<0$ and $q=-(2^7/3^3+\Ctt^2)<0$ then
$2^2p^3+3^3q^2=-2^{12}/3^3+3^3(2^7/3^3+\Ctt^2)^2>0$ 
for any $\ttilde\ge 0$. 
We can express the unique real root as 
\begin{equation}
\y^2=\frac{4}{3}\left\{1+2\cosh\left[\frac{1}{3}{\rm arcosh}\left(1+\frac{3^3\Ctt^2}{2^7}\right)\right]\right\}
\label{Cardano}  
 \end{equation} 
and $y=\pm\sqrt{y^2}$ with $y^2$ from Eq.~(\ref{Cardano}) can now be
plugged into Eqs.~(\ref{cfs}) to obtain $\w(y)$ and $\z(y)$ that are
required to solve the pair of quadratic equations (\ref{ansatz}). 
The four roots of the transcendental quartic are hence given by
\begin{align}
  \tlambda_1(\ttilde)&=\frac{\y}{2}\left(1+\sqrt{1-4\w(\y)/\y^2}\right),\nonumber\\
  \tlambda_2(\ttilde)&=\frac{\y}{2}\left(1-\sqrt{1-4\w(\y)/\y^2}\right),\nonumber\\
  \tlambda_3(\ttilde)&=-\frac{\y}{2}\left(1-\sqrt{1-4\z(\y)/\y^2}\right),\nonumber\\
  \tlambda_4(\ttilde)&=-\frac{\y}{2}\left(1+\sqrt{1-4\z(\y)/\y^2}\right).
\label{four}  
\end{align}
Moreover, we find $\w(\y)/\y^2=(1-2/\y^2-\Ctt/\y^3)/2$ and
$\z(\y)/\y^2=(1-2/ \y^2+\Ctt/\y^3)/2$. 
Since $\y>0$ while
$\tlambda\in[0,1)$, $\tlambda_2,\tlambda_3$ in Eq.~(\ref{four}) are
excluded automatically (note also that the square root in
$\tlambda_2,\tlambda_3$ becomes complex for $\ttilde\to\infty$). 
We also have $\lim_{\ttilde\to\infty}\y^2=4$ and $\lim_{\ttilde\to\infty}\w(\y)=1$ such
that $\lim_{\ttilde\to\infty}\tlambda_1=\tlambda_2=1$. 
Conversely, we find that
$\lim_{\ttilde\to 0}\ttilde^{2/3}\y^2=\C^{2/3}=\lim_{\ttilde\to 0}\ttilde^{2/3}\Ctt/\y$
such that $\lim_{\ttilde\to 0}\w(\y)=-1$ while $\lim_{\ttilde\to 0}\w(\y)\y^2=-{\Ctt}^{2/3}=0$. 
Therefore, $\lim_{\ttilde\to 0}\tlambda_1=\y\to\infty$ whereas 
$\lim_{\ttilde\to 0}\tlambda_2(\ttilde)=\y\times 0/2\searrow 0$. 
We recall that $\tlambda\in[0,1)$ which therefore excludes
$\tlambda_1(\ttilde)$ and identifies $\tlambda^{\dagger}(\ttilde)=\tlambda_2(\ttilde)$ as the supremum.
Finally, we introduce the auxiliary functions
\begin{equation}
g(u)\equiv\frac{2}{\sqrt{3}}\left\{1+2\cosh\left[\frac{1}{3}{\rm arcosh}\left(1+\frac{3^3}{2^7u^2}\right)\right]\right\}^{1/2}
\quad \text{and} \quad
\Lambda(u)\equiv
\frac{1}{2}\left[g(u)-\sqrt{4+2/g(u)u -g(u)^2}\right],
\end{equation}
as well as
\begin{equation}
h_-(u)\equiv \Lambda(u)u-\frac{1}{2}\frac{\Lambda(u)^2}{1-\Lambda(u)^2}
\end{equation}
which allows us to obtain and write the Cram\'er transform as
\begin{equation}
\phi_{\delta\tau}^*(\ttilde; \C)= \C h_-(\ttilde /\C) = \C h_-(\mu_1 t /\C),\quad 0\leq t\leq\mtau.
\label{SM_CT_L}
\end{equation}
In the last step we use Chernoff's inequality to obtain
the bound 
$\mathbbm{P}(\delta\ntau\geq t)\leq \e^{-n\phi^*_{\delta \tau}(t;\C)}\equiv\mathcal{U}^-_n(t;\C)$
for $0\leq t\leq\mtau$
which completes the proof of Theorem 2 and hence the derivation of the lower expression in Eq.~(7) of the Letter.

\begin{figure}[htbp]
\centering
\includegraphics[scale=1]{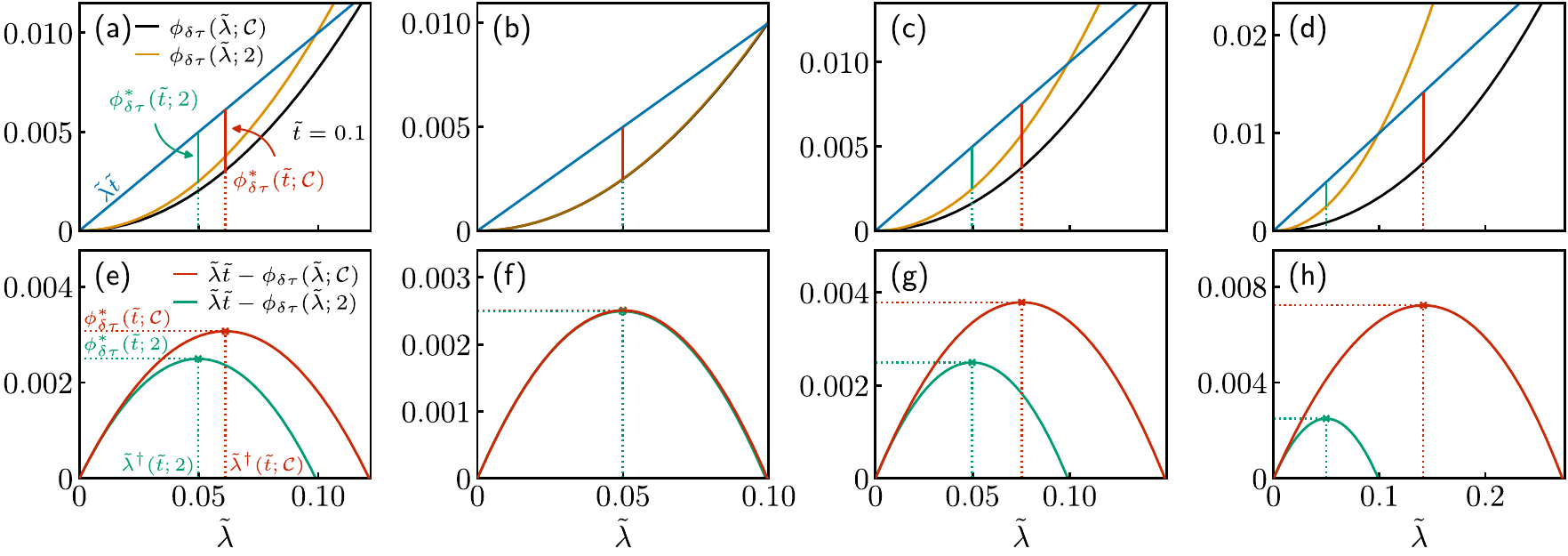}
\caption{Illustration of the Cram\'er-Chernoff bounding method for the left tail with 
$\ttilde=0.1$ and parameters for spatially confined Brownian search process in dimensions $d=1$ (a,e)
or $d=3$ (b,f), and discrete-state Markov jump processes
for the model of calmodulin (c,d) and a 8-state toy protein (d,h).
Top row depicts bounds of the cumulant generating function $\phi_{\delta\tau}(\tlambda;\C)$ (black)
and $\phi_{\delta\tau}(\tlambda;2)$ (yellow)
as a function of $\tlambda$, respectively.
Bottom row shows the differences $\tlambda\ttilde-\phi_{\delta\tau}(\tlambda;\C)$ (red)
and $\tlambda\ttilde-\phi_{\delta\tau}(\tlambda;2)$ (green) as a function of $\tlambda$, respectively (see also top row with $\tlambda\ttilde$ in blue).
The corresponding suprema are obtained at $\tlambda^\dagger(\ttilde;\C)$ and $\tlambda^\dagger(\ttilde;2)$ (dotted lines)
and define the Cram\'er transforms $\phi_{\delta\tau}^*(\ttilde;\C)$ and 
$\phi_{\delta\tau}^*(\ttilde;2)$ (compare top row).
For all considered models we demonstrate $\phi_{\delta\tau}(\tlambda;\C)\leq \phi_{\delta\tau}(\tlambda;2)$ and $\phi_{\delta\tau}^*(\ttilde;\C)\geq\phi_{\delta\tau}^*(\ttilde;2)$
as derived in the maintext.
Note for the panels (b,f) we have 
$\phi_{\delta\tau}(\tlambda;\C)\lessapprox\phi_{\delta\tau}(\tlambda;2)$
and $\phi_{\delta\tau}^*(\ttilde;\C)\gtrapprox\phi_{\delta\tau}^*(\ttilde;2)$
since $\C= 1.99\approx 2$.}
\label{fig:SM_CT_L}
\end{figure}

\subsection{Behavior of upper bounds $\mathcal{U}^\pm_n(t)$ for large sample sizes}
Here, we present some further remarks about the limit of large sample sizes.
Asymptotically as $n\to\infty$, $\mathcal{U}^\pm_n(t)$ is substantial only for $\tilde{t}/\C \ll 1$.
For the right tail bound $h_+(u)$
we immediately find that
for $u\ll 1$ we can Taylor expand $\sqrt{1+2u}=1+u-u^2/2+\mathcal{O}(u^3)$.
Consequently we directly obtain $h_+(u)=u^2/2 -\mathcal{O}(u^3)$, i.e., the upper tail
is sub-Gaussian
for small deviations and will converge to a Gaussian as $n\to\infty$.
For the left tail we furthermore have 
${\rm arcosh}(1+x)=\ln(1+x+\sqrt{x(x+2)})$ 
and thus $\lim_{x\to\infty}{\rm arcosh}(1+x)=\ln(2x)-1/(2x)^2$. 
As a result it follows that $\frac{1}{3}\lim_{u\to 0}{\rm arcosh}(1+3^3/2^7u^2)\simeq\frac{1}{3}\ln(3^3/2^6u^2)=\ln(3/4u^{2/3})-u^42^{12}/3^7$
and thus
\begin{align}
\lim_{u\to 0}g(u)&\simeq
\frac{2}{\sqrt{3}}\{1+2\cosh\ln(3/4u^{2/3})\}^{1/2}
=\frac{2}{\sqrt{3}}[1+3/4u^{2/3}]^{1/2}\\
&=u^{-1/3}[1+4u^{2/3}/3]^{1/2}
=u^{-1/3}[1+2u^{2/3}/3+\mathcal{O}(u^{4/3})]
=u^{-1/3}+\frac{2}{3}u^{1/3}+\mathcal{O}(u).
\end{align}  
A lengthy but straightforward calculation subsequently reveals that 
$\lim_{u\to 0}\Lambda(u)=u-\mathcal{O}(u^3)$ such that
\begin{align}
\lim_{u\to  0}\frac{\Lambda(u)^2}{1-\Lambda(u)^2}
\simeq\frac{u^2}{1-u^2}
=u^2+\mathcal{O}(u^4).
\end{align}
We therefore have that $\lim_{u\to 0}h_-(u)=u^2/2-\mathcal{O}(u^4)$, i.e., both
tails are sub-Gaussian for $\tilde{t}/\C \ll 1$
with $\C\equiv \mu_1^2 \langle\tau^2\rangle$.

\subsection{Proof of bounds on $\C$ and model-free concentration
  inequalities}
Notably, system details only enter
the Cram\'er transforms~\eqref{SM_CT_R} and~\eqref{SM_CT_L} 
(and consequently upper bounds on the deviation probability due to Chernoff's inequality) 
in the form of a \emph{system-specific constant}
$\C\equiv\mu_1^2\langle\tau^2\rangle$. Note that here we only allow
for equilibrium initial conditions.
Recalling that the moments of the first-passage time $\tau$ are expressed as 
$\langle \tau^m\rangle=m!\sum_{k>0}w_k/\mu_k^m$
allows us to write
\begin{equation}
0\leq2\frac{w_1}{\mu_1^2}
\leq
\langle\tau^2\rangle
=
2\sum_{k>0}\frac{w_k}{\mu_k^2}
\leq
2\sum_{k>0}\frac{w_k}{\mu_1^2}
=
\frac{2}{\mu_1^2}
\label{SM_2ndmoment_bound}
\end{equation}
where we have used that $w_k$ are non-negative, normalized,
and $\mu_1\leq\mu_{k>1}$.
Consequently, by Eq.~\eqref{SM_2ndmoment_bound}, 
we immediately find that the
system-constant itself is bounded 
$0\leq 2w_1\leq\C\leq 2$.
Note that analogous considerations can be used to
more generally 
obtain
$0\leq m! w_1\leq\mu_1^m\langle\tau^m\rangle\leq m!$
for the $m$-th moment, i.e., 
for $m=1$ we find the chain of inequalities $0\leq w_1\leq\mu_1\mtau\leq 1$.

The fact that $\C\in(0,2]$ can now be further leveraged to arrive at the model-free bounds
(Eq.~(11) in the Letter) which require no information about the underlying system.
Recall the upper bounds of the cumulant generating function 
$\phi_{\delta\tau}(\tlambda;\C)$ and their corresponding Cram\'er transform
$\phi_{\delta\tau}^*(\ttilde;\C)$, i.e., 
\begin{align}
\phi_{\delta\tau}(\tlambda;\C)=
\begin{dcases}
\frac{\tlambda^2}{2}\frac{\C}{1-\tlambda}&\tau\ge\mtau
\\
\frac{\tlambda^2}{2}\frac{\C}{1-\tlambda^2}&\tau<\mtau,
\end{dcases}
\qquad
\text{and}
\qquad
\phi_{\delta\tau}^*(\ttilde;\C)=
\begin{dcases}
\C h_+(\ttilde/\C) &\tau\ge\mtau
\\
\C h_-(\ttilde/\C) &\tau<\mtau.
\end{dcases}
\end{align}
Since $\phi_{\delta\tau}(\tlambda;\C)$ is monotonically increasing in $\C$ it follows that
$\phi_{\delta\tau}(\tlambda;\C)\leq\phi_{\delta\tau}(\tlambda;2)$, $\forall\tlambda\in[0,1)$
(see Figs~\ref{fig:SM_CT_R} and~\ref{fig:SM_CT_L} top row).
By definition of
$\phi_{\delta\tau}^*(\ttilde;\C)\equiv\sup_{\tlambda\in[0,1)}(\tlambda\ttilde-\phi_{\delta\tau}(\tlambda;\C))$
this bound in turn implies that $\phi_{\delta\tau}^*(\ttilde;\C)\geq\phi_{\delta\tau}^*(\ttilde;2)$
(compare Figs.~\ref{fig:SM_CT_R} and~\ref{fig:SM_CT_L} bottom row).
With Chernoff's inequality we moreover arrive at 
$\mathbbm{P}(\delta\ntau\geq t)\leq\e^{-n\phi^*_{\delta \tau}(t;\C)}\leq\e^{-n\phi^*_{\delta \tau}(t;2)}$
and hence
$\mathcal{U}^\pm_n(t;\C)\leq \mathcal{U}^\pm_n(t;2)$
which completes the derivation of Eq.~(11) in the Letter.

\newpage
\section{Model systems and details on numerical methods}
In the Letter we exemplify our results by 
considering
a Brownian molecular search process in dimensions $d=1$ and $d=3$, as well as 
discrete-state Markov-jump models of protein folding
for a 8-state toy protein and
the experimentally inferred model of calmodulin (compare Fig.~1b-d).
In this section we present further details on the model systems 
and their numerical treatment.

\subsection{Continuous-time discrete-state Markov jump process}
\label{sec:SM_MJP}
As illustrative discrete-state continuous-time
Markov-jump models of protein folding
we consider a simple 8-state toy protein~\cite{SM_bowman2013introduction,SM_Hartich2019}
and further use the experimentally inferred
folding network of the cellular calcium sensor protein calmodulin~\cite{SM_stigler2011complex}.
Since we consider \emph{equilibrium} initial conditions, proteins
start from an initial state drawn from the 
equilibrium density $\tilde{p}_{\rm eq}(x_0)$---note that the tilde denotes that the absorbing target is excluded---from
which they search the native state $a$ (here $a=(1,1,1)$ 
for the 8-state model and $a=F_{1234}$ for calmodulin; cf.~Fig.~1b-d).
Arrows in the networks denote possible transitions,
e.g.~a transition from state $i$ to state $j$ that occurs with the corresponding rate $L_{ji}$.
We consider \emph{reversible dynamics}, i.e., 
the resulting transition matrix $\hat{L}$ of the relaxation process
satisfies \emph{detailed balance} 
$p_{\rm eq, \textit{j}}/p_{\rm eq, \textit{i}}=L_{ji}/L_{ij} = \exp(F_i - F_j)$
and transitions rates are connected to 
the free energy of the states $F_i$~\cite{SM_Seifert2019}.

We recall that the first-passage time density $\wp_a(t)$ can 
be evaluated by using the spectral representation~\eqref{SM_FPT_density}.
To this end we set up the modified transition matrix, adopting in this section the Dirac bra-ket notation, 
$\hat{L}_a=\hat{L}- |a\rangle\langle a|$ 
where $|a\rangle\equiv(0,\ldots, 0,1,0,\ldots)^{\intercal}$ 
defines a vector with all entries zero expect at the $a$-th position of the absorbing state
where it equals one.
This effectively removes all transitions that
correspond to jumps leaving the absorbing state $a$.
Next, we carry out an eigendecomposition of $\hat{L}_a$
and determine the eigenvalues $\mu_k$, right eigenvectors $|\phi^{\rm R}\rangle$, and left eigenvectors $\langle\phi^{\rm L}|$.
We subsequently use obtained eigenmodes 
to compute the first-passage weights 
$w_k(x_0) = -\langle a|\phi_k^{\rm R}\rangle\langle\phi_k^{\rm L}|x_0\rangle$
(see~\cite{SM_Hartich2018, SM_Hartich2019}),
and recall that $\mu_k$ and $w_k$ determine 
the moments according to 
$\langle \tau^m\rangle=m!\sum_{k>0}w_k/\mu_k^m$.
Corresponding relevant parameters of the Markov jump models
are listed in Tab.~\ref{SM_Tab_MJP}.
Next we give further details on how matching transition rates 
are constructed.
\begin{table}[hbtp]
\caption{
Parameters for the Markov jump models for the 8-state toy protein and the inferred model of calmodulin.
Listed are values for the first-passage eigenvalues $\mu_k$, first-passage weights $w_k$, and 
the first $\mtau$ and second moment $\langle\tau^2\rangle$. 
}
\begin{tabular}{lcccccccccccccccc}  
\toprule
Model & $\mu_1$ & $w_1$ & $\mu_2$ & $w_2$ & $\mu_3$ & $w_3$ & $\mu_4$ & $w_4$ & $\mu_5$ & $w_5$ & $\mu_6$ & $w_6$ & $\mu_7$ & $w_7$ & $\mtau$ & $\langle\tau^2\rangle$\\ 
\midrule 
Toy protein & 0.976 & 0.337 & 6.148 & 0.009 & 1.551 & 0.583 & 4.203 & 0.001 & 4.396 & 0.0001 & 6.233 & 0.060 & 12.834 & 0.010 & 0.385 & 0.713 \\
Calmodulin  & 0.469 & 0.651 & 3.763 & 0.349 & 19.097 & 9.98E-5 & 143.749 & 2.42E-9 & 1581.629 & 1.52E-6& -- & -- & -- & -- & 1.479 & 5.958 \\
\bottomrule
\end{tabular}
\label{SM_Tab_MJP}
\end{table}

\subsubsection{Transitions rates of the 8-state toy protein model}
For the 8-state toy protein model 
we randomly generate a free energy level $F_i$ for each state $i\in\{1,2,3,4,5,6,7,8\}$ with $F_i$ 
uniformly distributed within the 
interval $0\leq F_i\leq 10$.
Transition rates that satisfy detailed balance are then obtained using the ansatz
\begin{equation}
	k_{i\to j}\equiv L_{ji} = \exp(\Delta F_i /2) \qquad \text{and} \qquad k_{j\to i}\equiv L_{ij} = \exp(-\Delta F_i /2),
\label{SM_Rate_Ansatz}
\end{equation}
where $\Delta F_i\equiv F_i-F_j$
and thus $\ln(L_{ji}/L_{ij}) = \Delta F_i = F_i - F_j$.
Obtained individual transition rates are listed in Tab.~\ref{SM_Tab_ToyModel}.
\begin{table}[hbtp]
\caption{Transition rates for the 8-state toy protein model obtained via the ansatz described in the main text.}
\begin{tabular}{cccccccc} 
\toprule
transition & rate $k_{i\to j}$ & transition & rate $k_{i\to j}$ & transition & rate $k_{i\to j}$ & transition & rate $k_{i\to j}$\\
\cmidrule[0.4pt](r{0.125em}){1-2}\cmidrule[0.4pt](lr{0.125em}){3-4}\cmidrule[0.4pt](lr{0.125em}){5-6}\cmidrule[0.4pt](l{0.125em}){7-8}
$1\to 2$ & 1.878 & $2\to 5$ & 2.648 & $3\to 7$ & 4.549 & $5\to 8$ & 0.106\\
$2\to 1$ & 5.327 & $5\to 2$ & 3.421 & $7\to 3$ & 1.00994 & $8\to 5$ & 124.477\\
\cmidrule[0.4pt](r{0.125em}){1-2}\cmidrule[0.4pt](lr{0.125em}){3-4}\cmidrule[0.4pt](lr{0.125em}){5-6}\cmidrule[0.4pt](l{0.125em}){7-8}
$1\to 3$ & 0.00463 & $2\to 6$ & 0.527 & $4\to 6$ & 0.358 & $6\to 8$ & 0.712\\
$3\to 1$ & 0.507 & $6\to 2$ & 36.0577 & $6\to 4$ & 36.457 & $8\to 6$ & 15.794\\
\cmidrule[0.4pt](r{0.125em}){1-2}\cmidrule[0.4pt](lr{0.125em}){3-4}\cmidrule[0.4pt](lr{0.125em}){5-6}\cmidrule[0.4pt](l{0.125em}){7-8}
$1\to 4$ & 0.326 & $3\to 5$ & 1.109 & $4\to 7$ & 0.523 & $7\to 8$ & 0.322\\
$4\to 1$ & 0.623 & $5\to 3$ & 0.0371 & $7\to 4$ & 6.670 & $8\to 7$ & 56.998\\
\bottomrule
\end{tabular}
\label{SM_Tab_ToyModel}
\end{table}

\subsubsection{Transitions rates of the calmodulin protein model}
In the experimental setup a constant external force $f$,
a so-called \emph{pretension},
is applied to the calmodulin protein via optical tweezers~\cite{SM_stigler2011complex}. 
Folding and unfolding processes are observed 
at different pretensions ranging from $6\,\text{pN}$ to $13\,\text{pN}$ and
corresponding force-dependent transition rates 
$k_{i\to j}(f)= L_{ji}(f)$ between two conformational states $i$ and $j$ are measured.
Note that $i,j \in \{\text{Unfold}, F_{12}, F_{123}, F_{23}, F_{34}, F_{1234} \}$
and we further map states according to 
$\text{Unfold}\leftrightarrow 1$,
$F_{12}\leftrightarrow 2$,
$F_{123}\leftrightarrow 3$,
$F_{23}\leftrightarrow 4$, 
$F_{34}\leftrightarrow 5$,
and
$F_{1234}\leftrightarrow 6$
for convenience.
For our purposes we choose, without loss of generality, a pretension of $f=9\,\text{pN}$
and obtain the corresponding measured transitions rates from Fig.~S8 
in the Supplementary Material of~\cite{SM_stigler2011complex}.
Clearly, experimental transitions rates
are accompanied with measurement uncertainties which is reflected in 
slight ``deviations'' from a mathematically precise definition of 
detailed balance.
To mitigate this issue, and to ensure that transition rates precisely obey 
detailed balance 
$k_{i\to j} p_{\rm eq, \textit{i}}= k_{j\to i} p_{\rm eq, \textit{j}}$,
we further have to slightly adjust the rates.

First, we compute the invariant density $p_{\rm eq}$ from the 
experimental rates and obtain a corresponding 
free energy level $F_i = -\ln(p_{\rm eq, \textit{i}})$.
Next, we use the ansatz~\eqref{SM_Rate_Ansatz}, i.e.,
$L_{ji} = A_i \exp(\Delta F_i /2)$ 
and $L_{ij} = A_i \exp(-\Delta F_i /2)$
where we introduce a constant $A_i$.
Finally, $A_i$'s are chosen such that resulting 
transition rates fall within experimental error bars in Ref.~\cite{SM_stigler2011complex}.
Obtained transition rates are listed in Table~\ref{SM_Rate_Cal}.
\begin{table}[hbtp]
\caption{Transition rates of the Markov jump model for the calmodulin protein.
Rates are extracted from the Supplemental Material of Ref.~\cite{SM_stigler2011complex}
and modified such that they obey detailed balance precisely according to the maintext.}
\begin{tabular}{cccccc} 
\toprule
transition & rate $k_{i\to j}$ & transition & rate $k_{i\to j}$ & transition & rate $k_{i\to j}$\\
\cmidrule[0.4pt](r{0.125em}){1-2}\cmidrule[0.4pt](lr{0.125em}){3-4}\cmidrule[0.4pt](l{0.125em}){5-6}
$1\to 2$ & 5.997 & $1\to 4$ & 13.439 & $1\to 5$ & 15.330\\
$2\to 1$ & 0.774 & $4\to 1$ & 127.968 & $5\to 1$ & 0.121\\
\cmidrule[0.4pt](r{0.125em}){1-2}\cmidrule[0.4pt](lr{0.125em}){3-4}\cmidrule[0.4pt](l{0.125em}){5-6}
$5\to 6$ & 3.749 & $2\to 3$ & 1514.820 & $2\to 6$ & 13.441\\
$6\to 5$ & 13.326 & $3\to 2$ & 53.0661 & $6\to 2$ & 2.922\\
\bottomrule
\end{tabular}
\label{SM_Rate_Cal}
\end{table}
\vspace{-0.5cm}
\subsection{Spatially confined Brownian molecular search process}
\label{SM_SecBessel}
We also test our theory for Markov processes on a continuous state-space.
More precisely, we consider the spatially confined 
diffusive search of a Brownian particle 
in a $d$-dimensional unit sphere with a reflecting boundary at $R=1$
and a perfectly absorbing spherical target of radius $0<a<1$, here $a=0.1$, in the
center (compare Fig.~1b).
The closest distance of the particle to the surface of the 
absorbing sphere at time $t$ is a confined Bessel process (see e.g.~\cite{SM_Pitman1975,SM_Barkai2014,SM_Hartich2019})
which time evolution obeys the It\^{o} equation
\begin{equation}
	dx_t=(d-1)x_t^{-1}dt+\sqrt{2}dW_t,
\label{SM_Bessel}
\end{equation}
where $dW_t$ is the increment of a Wiener process (i.e.~Gaussian white noise) with
$\langle dW_t\rangle=0$ and $\langle
dW_tdW_{t'}\rangle=\delta(t-t')dt$, and we have set, without loss of
generality, $D=1$.
The general case with any $0<D<\infty$ and a sphere of radius
$R$ is covered by expressing
time in units of $R^2/D$. 

For $d=1$ Eq.~\eqref{SM_Bessel} reduces to a
$1$ dimensional Brownian motion
which has the equilibrium first-passage weights
\begin{equation}
w_k^{\rm eq}=
\frac{2}{\pi^2}\frac{1-\sin\left[(k-1)\pi\right]}{(k-1/2)^2}
\end{equation}
and matching first-passage eigenvalues are obtained as $\mu_k=\pi^2 (k-1/2)^2$.
Moreover, for $d=3$ the first-passage time probability density 
of the Bessel process 
can be evaluated exactly
and has the equilibrium weights
\begin{equation}
w_k^{\rm eq}=\frac{2}{\mu_k}\frac{3a^2}{1-a^3}\frac{\tan[(1-a)\sqrt{\mu_k}]+\frac{1}{\sqrt{\mu_k}}}{(1-a)\tan[(1-a)\sqrt{\mu_k}]-\frac{a}{\sqrt{\mu_k}}},
\label{Bessel}  
\end{equation}
with the first-passage eigenvalues $\mu_k$ being the solutions of the transcendental equation
$\sqrt{\mu_k}=\tan([1-a]\sqrt{\mu_k})$ that can be solved 
analytically using Newton's series \cite{SM_Hartich2019}.
Relevant parameters for the spatially confined Brownian search process with $a=0.1$ are listed in Tab.~\ref{SM_Tab_Bessel}.

{Note that the mean for $d=3$ and arbitrary $x_0$ can be obtained directly~\cite{SM_godec2017first}
as $\langle\tau_a(x_0)\rangle = [1/a -1/x + a^2/2 -x^2/2]/3$ which integrated over all starting positions, i.e.~equilibrium initial conditions, 
gives the global mean first-passage time
\begin{equation}
\langle\tau_a^{\rm eq}\rangle =
\left[ \frac{4\pi}{3}(1-a^3) \right]^{-1}
\int_a^1  \frac{4\pi}{3}\left[\frac{1}{a} - \frac{1}{x} + \frac{a^2}{2} -\frac{x^2}{2} \right] dx
=
\frac{1}{1-a^3}\left[\frac{1-a^3}{3a} - \frac{1-a^2}{2} + \frac{a^2(1-a^3)}{6} - \frac{1-a^5}{10}\right].
\end{equation}
}

\begin{table}[hbtp]
\caption{
Parameters for the spatially confined Brownian molecular search process in dimensions $1$ and $3$.
Listed are values for the
first $5$ first-passage eigenvalues $\mu_k$, first-passage weights $w_k$, and 
the first $\mtau$ and 
second moment 
$\langle\tau^2\rangle$, respectively~\footnote{
For the numerical evaluation of $\mtau$ and $\langle\tau^2\rangle$ as listed we truncate the 
sum after $M=1000$ terms.}. 
}
\begin{tabular}{lcccccccccccc} 
\toprule
Model & $\mu_1$ & $w_1$ & $\mu_2$ & $w_2$ & $\mu_3$ & $w_3$ & $\mu_4$ & $w_4$ & $\mu_5$ & $w_5$ & $\mtau$ & $\langle\tau^2\rangle$\\ 
\midrule 
1D Brownian motion & 2.467 & 0.811 & 22.207 & 0.0901 & 61.685 & 0.0324 & 120.903 & 0.0165 & 199.859 & 0.01001 & 0.333 & 0.267\\
3D Bessel process & 0.363 & 0.994 & 25.174 & 0.00277 & 73.926 & 9.163E-4 & 147.037 & 4.573E-4 & 244.516 & 2.742E-4 & 2.739 & 15.09\\
\bottomrule
\label{SM_Tab_Bessel}
\end{tabular}
\end{table}

\newpage
\subsection{Statistics of first-passage times $\tau$, the sample mean $\ntau$, maximum $\tau^+_n$, and minimum $\tau^-_n$}
\label{SM_Sampling}
Finally we provide some further details on the sampling method used to 
obtain the statistics of (i) the first-passage time $\tau$,
(ii) the sample-mean $\ntau\equiv\sum_i \tau_i/n$, and
{(iii) the expected maximal and minimal deviation from the mean $\langle \tau^\pm_n - \mtau \rangle$
for a sample with a fixed number $n$ of independent realizations for all considered models. 
}

\begin{figure}[htbp]
\centering
\includegraphics[scale=1]{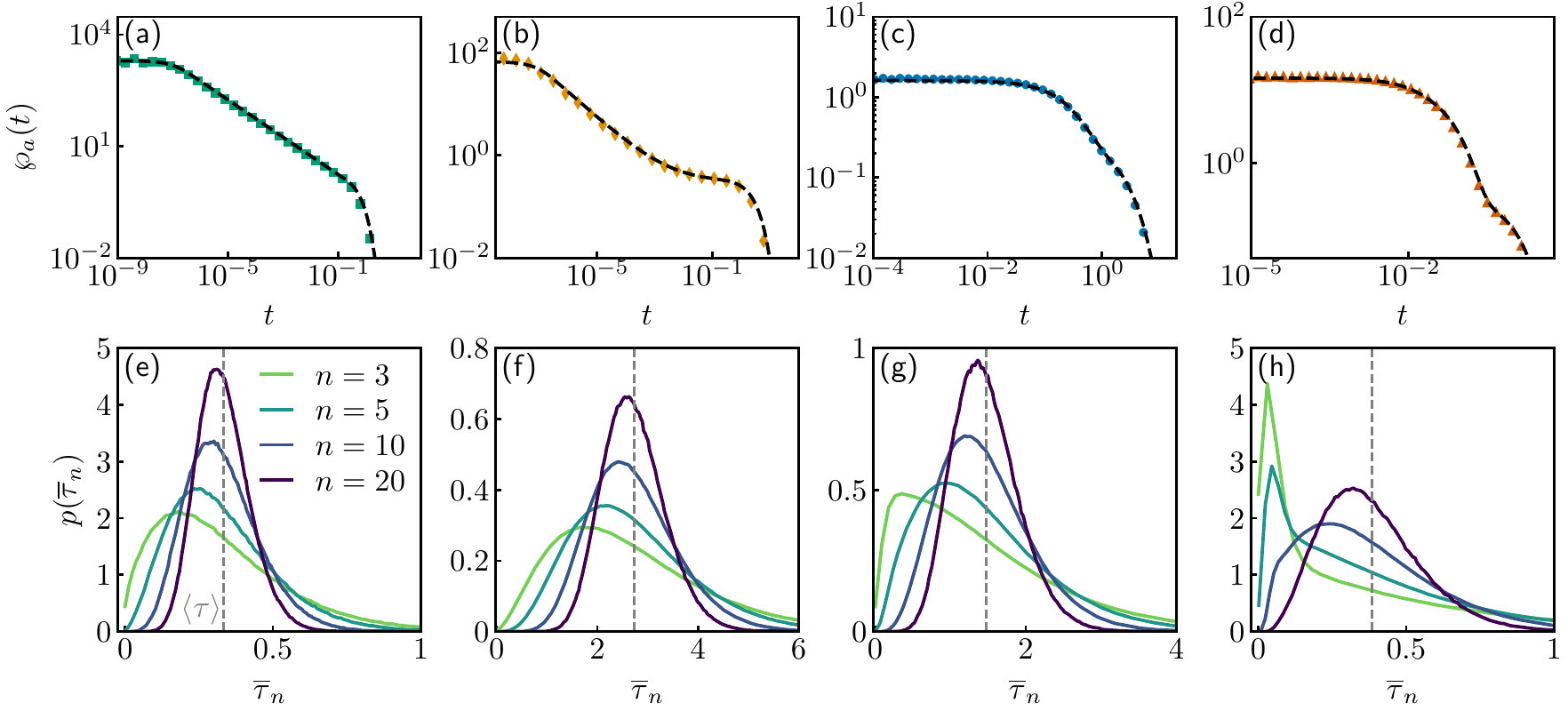}
\caption{Inversion sampling of first-passage statistics 
for a confined Brownian search process in dimensions (a,e) $d=1$
and (b,f) $d=3$, and discrete-state Markov jump processes
for (c,d) the calmodulin network and (d,h) a 8-state toy protein.
(a-d) First-passage time density $\wp_a(t)$ obtained using inversion sampling (symbols) and 
analytical result as black dashed lines.
(e-h) Empirical probability density of the sample mean $\ntau$ for different $n$ values. 
True mean first-passage times $\mtau$ are shown in grey.
}
\label{fig:SM_Sampling}
\end{figure}

We recall that after determining the first-passage eigenvalues $\mu_k$ and
first-passage weights $w_k$,
the first-passage time density 
$\wp_a(t)$~\eqref{SM_FPT_density}
and survival probability 
$S_a(t)$~\eqref{SM_S_Prob} are fully characterized.
To now sample the random variable $\tau$, i.e.~individual realizations of the first-passage process, we 
employ the so-called \emph{inversion sampling method}~\cite{SM_Devroye1986}.
This method allows us to generate independent samples 
of $\tau$ from
$\wp_a(t)$ given its cumulative distribution function (CDF) 
which is directly related to the survival probability according to $1-S_a(t)$.
Note that for discrete-state dynamics 
the number of states $M$ is finite, i.e.~$k=1, \ldots, M$,
and therefore Eq.~\eqref{SM_S_Prob} (and hence the CDF) is a finite sum.
In contrast, for continuous-state dynamics we formally have $M=\infty$, meaning that sums are here not finite.
For the following numerical evaluation of the spatially confined
Brownian search process we therefore truncate the 
sum after $M=1000$ terms.
The first-passage time densities $\wp_a(t)$ obtained via inversion sampling (symbols) for all considered models
are shown in Fig.~\ref{fig:SM_Sampling}a-d
and corroborated by the corresponding analytical result~\eqref{SM_FPT_density} (dashed black line).

For Fig.~2a-d in the Letter 
empirical
probabilities that
${\ntau-\langle\tau\rangle}$ lies within a desired range of $\pm$ 10\% of the
longest first-passage time scale
$\mu_1^{-1}$, $\mathbbm{P}(\mu_1[\ntau-\langle\tau\rangle]\in
     [-0.1,0.1])$,
are computed using statistics of the sample mean $\ntau$
by fixing $n$, i.e., the number of individual realizations the average is taken over.
In particular, we have $n\in\{1,2,3,5,10,20,30,40,50,75,100,150,200,300,400,500\}$.
Subsequently, for each individual fixed $n$ the sample mean $\ntau$ itself
is sampled a total of $N=10^6$ times.
That is, we first draw $n$ first-passage times $\tau$, compute $\ntau$ by averaging over the drawn $n$ realizations, 
and finally repeat this step $N=10^6$ times to obtain statistics of $\ntau$
for all $n$ values introduced above.
Probability densities of the sample mean are shown 
in Fig.~\ref{fig:SM_Sampling}e-h for $n\in\{3,5,10,20\}$ and all model systems.
Corresponding true mean first-passage times $\mtau$ are highlighted in grey.

In Fig.~2e-h of the Letter the
probabilities to deviate more than $t$ in either direction, $\mathbbm{P}(\pm[\ntau-\mtau]\geq t)$,
are computed from analogous statistics of the sample mean $\ntau$. 
Since we also consider empirical probabilities
for rare events with large deviations (i.e.~large $\mu_1 t$) we however require 
substantially more statistics of $\ntau$. 
To this end we now have 
$N=10^7$ for $n\in\{1,3\}$ and $N=10^{11}$ for $n\in\{5,10,20\}$.
In addition it should be further noted that we re-scale obtained probabilities according to $\mathbbm{P}^{1/n}$.
To compute an empirical deviation probability
where e.g.~$\mathbbm{P}^{1/20} = 0.1$ one would be thus required
to sample rare events that occur with a probability
of $\simeq 10^{-20}$.

In Fig.~3a of the Letter each data point corresponds to the relative 
error $\mu_1(\ntau-\mtau)$ 
(note that $\mu_1$ and $\mtau$ are different for each model)
where the sample mean $\ntau$ is again obtained by first
fixing $n$ and then sampling $n$ first-passage times $\tau$ according to the inversion sampling method and subsequently taking the average.

{In Fig.~3d-e in the Letter and Figs.~\ref{fig:SM_Maxdev}
and~\ref{fig:SM_Mindev} below we show the expected deviation
of the maximum $\tau^+_n$ and minimum $\tau^-_n$ in a sample of $n$ realizations
from the mean first-passage time $\mtau$ as a function of $\mu_1\mtau$.
To sample values over the entire range of possible values $\mu_1
\mtau\in[0,1]$  we generate $10000$ random folding landscapes for the
Markov jump process network of Calmodulin and the 8-state toy protein
model, respectively.
Each data point shown corresponds to one particular realization of the Markov jump process for the given network 
topology 
where  transitions rates obeying detailed balance are constructed using the
methods described in Sec.~\ref{sec:SM_MJP}. 
The free energy levels $F_i$ are now uniformly distributed within the interval
$0\leq F_i\leq 20$.
For each random realization of a folding landscape (i.e., data point) we draw a
fixed number of independent first-passage times $n\in\{3,5,10,20\}$
using inversion sampling 
to determine the minimum $\tau^-_n$ and maximum $\tau^+_n$, receptively.
To compute the associated averages $\langle\tau^\pm_n\rangle$
we acquire minimal and maximal $\tau^\pm_n$'s for a total of $10^7$ independent draws of $n$ realizations for each of the $10000$ respective
random folding landscapes.
The same procedure is applied to the confined diffusion process where for $d=3$
we vary the target radius $a\in[0.05, 0.95]$.
Note that in the case of $d=1$ we have no parameter to vary, i.e.,
there is only one data point.}

\newpage
\section{Uncertainty quantification with confidence intervals}
\begin{figure}[hbtp]
\centering
\includegraphics{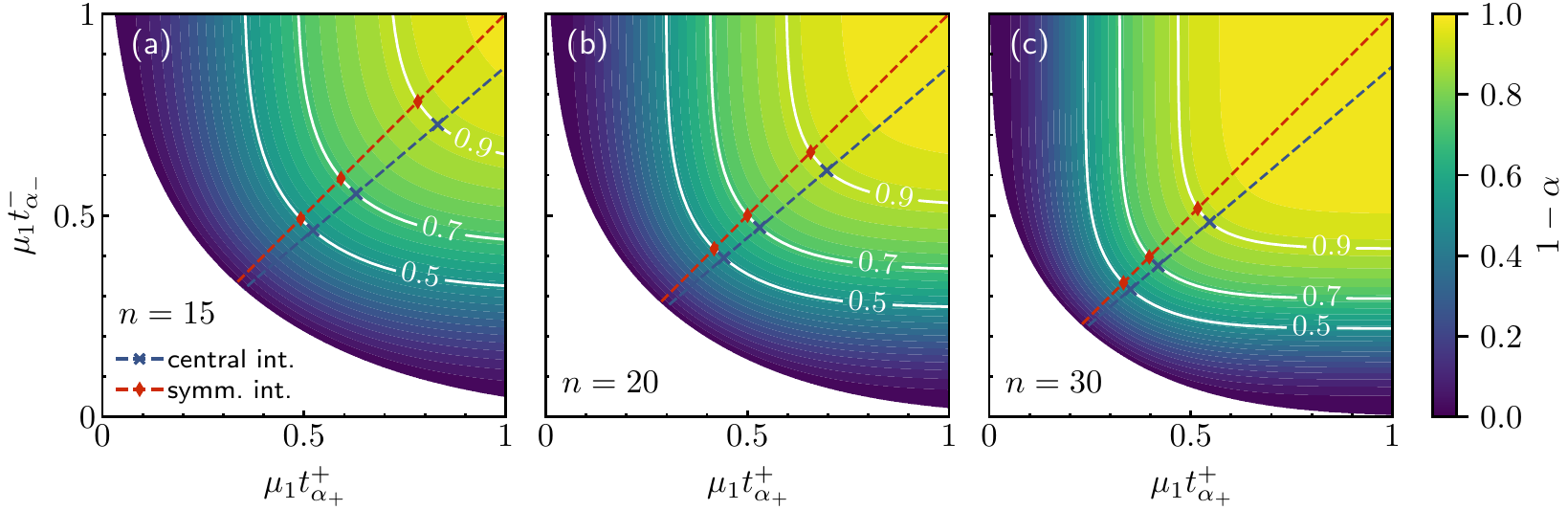}
\caption{
Contour plot of different choices of possible two-sided confidence intervals $[-\mu_1\tm, \mu_1\tp]$ for a fixed confidence level $\alpha$
and (a) $n=15$, (b) $n=20$, (c) $n=30$.
Contour lines for $\alpha\in\{0.1,0.3,0.5\}$ are depicted in white.
Specific choices of central and symmetric are shown in blue and red, respectively,
and we let $\C=1$ for all panels.}
\label{fig:SM_ConfInt}
\end{figure}

In this section we extend the discussion and present some further details on the confidence intervals introduced in the Letter.
Our derived upper bounds $\mathcal{U}^\pm_n(t)$ 
can be applied to construct 
non-asymptotic performance guarantees
such as confidence intervals.
In particular, they can be employed to bound the 
probability that $\delta\ntau\equiv\ntau -\mtau$
is found to be in some interval $[-\tm, \tp]$, i.e., 
\begin{align}
\mathbbm{P}(\delta\ntau\in[-\tm, \tp])\nonumber
&=\mathbbm{P}(-\tm\leq\delta\ntau\leq\tp)\nonumber\\
&=\mathbbm{P}(\delta\ntau\geq -\tm \cap \delta\ntau\leq\tp)\nonumber\\
&\geq 1-\mathbbm{P}(\delta\ntau\leq -\tm)-\mathbbm{P}(\delta\ntau\geq \tp)\nonumber\\
&\geq 1-\underbrace{\mathcal{U}^-_n(\tm)}_{\equiv \alpha_-}-\underbrace{\mathcal{U}^+_n(\tp)}_{\equiv\alpha_+}.
\label{SM_Conf_Int}
\end{align}
In passing from the second to the third line we have applied 
Boole's second inequality, and from the third to forth line we use bounds (7) of the Letter. 
In the last line we additionally introduced acceptable 
right and left tail error probabilities $\alpha_\pm$.
The implicit interval $[-\tm, \tp]$ therefore defines a confidence interval
at a confidence level of $1-\alpha$ with 
$\alpha\equiv\alpha_+ + \alpha_-$, and $\alpha_+ + \alpha_- < 1$.
In general the choice of the confidence interval for a fixed probability $1-\alpha$ is \emph{not} unique.
Some common options in the literature (see e.g.~\cite{SM_cowan1998statistical,SM_Lista2017}), 
all having the same confidence level, are listed below.
\begin{itemize}
	\item One common choice are so-called \emph{central} intervals (blue lines in Fig.~\ref{fig:SM_ConfInt}) which correspond to equal tail probabilities
	$\alpha_+=\alpha_-=\alpha/2$ for the complementary intervals $[-\mtau, -t^-_{\alpha_-}]$ and
	$[t^+_{\alpha_+},\infty)$.
	Notably, we remark that central confidence intervals
	do \emph{not} generally imply that $t^+_{\alpha_+}$ and $t^-_{\alpha_-}$
	are equidistant from another, i.e., $t^+_{\alpha_+}\neq t^-_{\alpha_-}$.
	\item As an alternative one could likewise choose 
	$t^+_{\alpha_+} =  t^-_{\alpha_-} \equiv \Delta t /2$, which subsequently	
	leads to the \emph{symmetric} interval $[-\Delta t/2, \Delta t/2]$
	with total length $\Delta t$ (see red lines in Fig.~\ref{fig:SM_ConfInt}) .
	Analogously, a symmetric interval does \emph{not} necessarily imply that the 
	corresponding tail probabilities are equal, i.e., in general
	$\alpha_+\neq\alpha_-$.
	\item Both considerations above lead to two-sided intervals.
	However, another possible choice includes the fully asymmetric intervals 
	$[-\mtau, \tp]$ and
	$[-\tm,\infty)$, i.e., \emph{one-sided} intervals 
	with a	corresponding confidence level $1-\alpha_+$ (for the upper limit $\tp$) 
	and $1-\alpha_-$ (for the lower limit $\tm$), respectively,
	\begin{equation}
		\mathbbm{P}(\pm\delta\ntau\leq t^{\pm}_{\alpha_\pm}) \geq 1-\alpha_\pm.
		\label{SM_ConfIntOneSide}
	\end{equation}
\end{itemize}

Confidence intervals are practically useful as they
answer questions such as e.g.:\\\\
\emph{How many realizations are required 
to achieve a desired accuracy with a specified probability?}\\
Or: \emph{For a given number of realizations a desired accuracy is achieved
with at least what probability?}\\

In the case of symmetric confidence intervals $\tp=\tm$ (see Fig.~\ref{fig:SM_ConfInt} red lines)
the interval endpoints are implicitly defined via the last line of Eq.~\eqref{SM_Conf_Int}
which is easily solved using standard root-finding procedures
like the bi-section method~\cite{SM_burden2015numerical}.
The same holds true for other interval choices, however, when
specifying the error probabilities $\alpha_\pm$ directly---as done 
for e.g.~two-sided central intervals ($\alpha_\pm = \alpha/2$) or one-sided intervals---it suffices to solve 
Eq.~\eqref{SM_ConfIntOneSide} with the respective $\alpha_\pm$.
Hereby, the lower confidence limit $\tm$ is again easily obtained using standard 
root-finding methods.
Notably, the upper confidence limit $\tp$
can now be solved analytically.
To show this we consider $\mathcal{U}_n^+(\tp;\mathcal{C}) =\alpha_+$, i.e., 
we identify the $\tp$ that solves
\begin{equation}
0=-n\C h_+(\mu_1 \tp / \C) - \ln(\alpha_+).
\end{equation}
The roots are identified as
\begin{align}
	t_1 = - \frac{\ln{\left(\alpha_+ \right)}}{\mu_1 n} - \frac{\sqrt{2} \sqrt{- \ln{\left(\alpha_+ \right)}}}{\mu_1 \sqrt{n/\C}}
	\qquad \text{and} \qquad
	t_2 = - \frac{\ln{\left(\alpha_+ \right)}}{\mu_1 n} + \frac{\sqrt{2} \sqrt{- \ln{\left(\alpha_+ \right)}}}{\mu_1 \sqrt{n/\C}},
\end{align}
and we identify $\tp=t_2$ as the relevant solution.
Having obtained an explicit expression for $\tp$ further allows us 
to re-insert it into the left-hand side of Eq.~\eqref{SM_ConfIntOneSide}, i.e., 
we find that with a probability of at least $1-\alpha_+$
\begin{equation}
\delta\ntau\leq- \frac{\ln{\left(\alpha_+ \right)}}{\mu_1 n} + \frac{\sqrt{2} \sqrt{- \ln{\left(\alpha_+ \right)}}}{\mu_1 \sqrt{n/\C}}.
\end{equation}

The required number of realizations $n^*$ 
to ensure with a probability of at least $1-\alpha$ that 
$\delta\ntau$ is found within some interval $[-\tm, \tp]$ 
(e.g.~symmetric interval in Fig.~3b) is analogous identified 
according to Eq.~(13) in the Letter
\begin{equation}
\mathcal{U}_{n^*}(\tp;\C) + \mathcal{U}_{n^*}(\tm;\C) = \alpha, 
\end{equation}
which once again is readily solved via e.g.~the bisection method.
Moreover, in the case of one-sided intervals one immediately finds
the corresponding analytical expression
\begin{equation}
n^* \geq -\frac{\ln(\alpha_\pm)}{\C h_\pm(\mu_1 t/\C)},
\end{equation}
where $n^*$ denotes the required number to ensure that 
$\pm\delta\ntau\leq t$ with at least $1-\alpha_\pm$.

\section{Controlling uncertainty of first-passage times when $\mtau$
  is an insufficient statistic}
Note
that whenever $\mu_1\mtau$ substantially differs from 1, or
essentially equivalently $\mathcal{C}$ becomes substantially smaller
than 2, many time-scales enter the problem and $\wp(t)$ tends to
largely differ from an exponential. In turn, $\mtau$ \emph{a priori}
becomes an insufficient statistic. However, as we prove below the
expected range of inferred $\tau$ in general, and thus even in cases
when $\mtau$ is an insufficient statistic, turns out to be sharply bounded from
above and below by functions of $\mu_1\mtau$ only. We therefore ultimately seek for bounds on
$\mu_1(\langle\tau^\pm_n-\mtau)$ as a function of
$\mu_1\mtau$. Therefore, even when $\mtau$ is  an \emph{a priori}
insufficient statistic it is useful and insightful to infer the
empirical first-passage time $\ntau$
and control its uncertainty, as it in turn sets sharp upper bounds on
the range of inferred first passage times.

\section{Proof of bounds on extreme deviations from $\mtau$}
In this section we prove lower ($\underline{\mathcal{M}}_n^\pm$) 
and upper ($\widebar{\mathcal{M}}^\pm_n$) bounds on
the expected deviation  of the maximum and minimum from the mean 
in a sample of $n$ i.i.d.~realizations of $\tau$. That is, 
we consider the average $\langle m_n^\pm\rangle$
of the random quantity $m_n^\pm \equiv \tau^\pm_n -\mtau$
where $\tau^+_n\equiv\max_{i\in[1,n]}\tau_i$ 
and $\tau^-_n\equiv\min_{i\in[1,n]}\tau_i$, respectively.
Note that here from we drop the subscript $a$ in the survival probability $S_a(t)\rightarrow S(t)$ for ease of notation.

Let $g_n^\pm(t)\equiv\frac{d}{dt}\mathbbm{P}(\tau^\pm_n\le
t)=-\frac{d}{dt}\mathbbm{P}(\tau^\pm_n> t)$ denote the probability
density of $\tau^\pm_n$.  
Since we consider positive random variables $\tau_i\geq 0$ we have
\begin{align}
  \langle\tau^\pm_n\rangle
  = \int_0^\infty t g^\pm_n(t)dt
  = \lim_{c\to\infty}\int_0^c t \left(-\frac{d}{dt}\mathbbm{P}(\tau^\pm_n> t)\right)dt 
  \overset{\rm P.I.}{=} \lim_{c\to\infty}\left[-t\mathbbm{P}(\tau^\pm_n> t)|_0^c+\int_0^c\mathbbm{P}(\tau^\pm_n> t)dt \right],
\end{align}
where in the last step we performed a partial integration. Because
$\mathbbm{P}(\tau\le 0)=0$ (while in general $\wp(0)\ge 0$), we may
further write
\begin{align}
&t\mathbbm{P}(\tau^\pm_n> t)\Big |_0^c=t(1-\mathbbm{P}(\tau^\pm_n\le
  t))\Big |_0^c=c\mathbbm{P}(\tau^\pm_n> c)=c\int_c^\infty g^\pm_n(t)dt\le \int_c^\infty
  tg^\pm_n(t)dt.
\end{align}
Now, since $\int_0^\infty t g^\pm_n(t)\le\infty,\forall n<\infty$, 
we have $\lim_{c\to\infty}\int_c^\infty tg^\pm_n(t)dt=0$ 
and in turn
\begin{align}
\langle\tau^\pm_n\rangle=\int_0^\infty\mathbbm{P}(\tau^\pm_n > t)dt.
\label{eq:SM_MinMaxMean}
\end{align}  
In addition we have 
$\mathbbm{P}(\tau^+_n>t)=1-\mathbbm{P}(\tau^+_n\le
t)=1-\mathbbm{P}(\tau\le t)^n
= 1-(1-S(t))^n$, as well as
$\mathbbm{P}(\tau^-_n >t) = \mathbbm{P}(\tau > t)^n = S(t)^n$, 
since $\tau_i$'s are i.i.d.~[see Eq.~\eqref{SM_EV}].
Therefore, according to Eq.~\eqref{eq:SM_MinMaxMean}, we find
\begin{align}
&\langle\tau^+_n\rangle 
= \int_0^\infty [1-(1-S(t))^n] dt 
= \int_0^\infty \left(1- \left[1-\sum_{k>0} w_k \e^{-\mu_k
    t}\right]^n\right) dt
\label{MAXX}
\\
&\langle\tau^-_n\rangle
= \int_0^\infty S(t)^n dt 
= \int_0^\infty \left[\sum_{k>0} w_k \e^{-\mu_k t}\right]^n dt,
\label{eq:SM_MaxMin}
\end{align}
where in the last step we used the spectral representation of $S(t)$
[Eq.~\eqref{SM_S_Prob}].

\subsection{Proof of upper bound on $\langle m_n^+\rangle$}
\label{sec:SM_uppermax}
We now prove the upper bound on the expected maximal deviation from the mean in sample of $n$ realizations
and we therefore examine $\langle m_n^+\rangle$ at any given fixed
value of $\mu_1\mtau$.

Because the first-passage density at equilibrium $\wp(t)$ is monotonically
decaying with $t$ and normalized (i.e.\ all $w_k\ge 0$ and $\sum_kw_k=1$; see Eq.~\eqref{eweights}),
the maximum $\langle \tau_n^+\rangle$ will obviously be maximized at fixed
$\mu_1\mtau$ when $w_1$---the weight of the longest first-passage
time-scale---is maximal (see Eq.~\eqref{MAXX}). 

As a first step we therefore consider how the constraint $\mu_1\mtau={\rm
  const.}$ (note that $\mu_1\mtau\in(0,1]$) affects the first-passage
        spectrum $\{w_k,\mu_k\}$, i.e.\
\begin{equation}
\mu_1\mtau = w_1 + \sum_{k\geq 2} w_k \frac{\mu_1}{\mu_k} = {\rm const.}, 
\end{equation}
which consequently implies, since all terms are positive, 
\begin{equation}
\sup\{w_1 : \mu_1\mtau = {\rm const.} \} = \mu_1\mtau,
\label{eq:SM_w1max}
\end{equation}
and in turn implies that a spectral gap $\frac{\mu_1}{\mu_k}\to 0$ for all
$k > 1$ maximizes $w_1$. 
Eq.~\eqref{eq:SM_w1max} therefore implies
\begin{equation}
\sup_{\{\mu_i, w_i\}|\mu_1\mtau={\rm const.}}\lim_{t\to\infty} S(t) = \mu_1 \mtau \e^{-\mu_1 t},
\end{equation}
such that we may use the maximizing ansatz
\begin{equation}
S_{\rm max}(t) = \lim_{\epsilon\to 0 } (1-\mu_1\mtau)\e^{-t/\epsilon} + \mu_1\mtau \e^{-\mu_1 t},
\label{eq:SM_Ansatz}
\end{equation}
where the first term embodies the spectral gap and is defined as a distribution. 
With the above ansatz 
$\langle \tau^+_n\rangle$ is maximal 
(and $\langle\tau^-_n\rangle$ is minimal; see
Sec.~\ref{sec:SM_minlower} for the corresponding lower bound) under the
constraint $\mu_1\mtau={\rm const.}$. 
We thus continue with Eq.~\eqref{eq:SM_Ansatz} to prove the upper bound on $\langle\tau^+_n\rangle$.
Note that in the following calculation
we omit $\lim_{\epsilon\to 0}$ for the moment and later explicitly
state when we take the limit. 
First we may write,
\begin{align}
S_{\rm max}(t)^m 
= \left[(1-\mu_1\mtau)\e^{-t/\epsilon} + -\mu_1\mtau \e^{-\mu_1 t} \right]^m
= \sum_{l=0}^m \binom{m}{l}(1-\mu_1\mtau)^{(m-l)} \e^{(m-l)t/\epsilon} (\mu_1\mtau)^l \e^{-\mu_1 l t}.
\label{eq:SM_Smax}
\end{align}
Furthermore, according to the binomial theorem $(1-S(t))^n = \sum_{k=1}^n(-1)^k \binom{n}{k} S(t)^n$ 
and using Eq.~\eqref{eq:SM_Smax} we find
\begin{equation}
(1-S_{\rm max}(t))^n  = \sum_{k=0}^n \binom{n}{k} (-1)^k \sum_{l=0}^k \binom{k}{l} (1-\mu_1\mtau)^{k-l} \e^{-(k-l)t/\epsilon}, (\mu_1\mtau)^l \e^{-\mu l t}
\end{equation}
and therefore also
\begin{align}
1-(1-S_{\rm max}(t))^n
=
-\sum_{k=1}^n \binom{n}{k} (-1)^k \sum_{l=0}^k 
\binom{k}{l} (1-\mu_1\mtau)^{k-l} \e^{-(k-l)t/\epsilon} (\mu_1\mtau)^l \e^{-\mu l t}.
\end{align}
We recall that the ansatz Eq.~\eqref{eq:SM_Ansatz} maximizes
$\langle\tau^+_n\rangle$, that is, it yields the upper bound [compare Eq.~\eqref{eq:SM_MaxMin}]
\begin{equation}
\langle \tau^+_n\rangle 
\leq \sum_{k=1}^n \binom{n}{k} (-1)^{k+1} \sum_{l=0}^k \binom{k}{l} (1-\mu_1\mtau)^{k-l} (\mu_1\mtau)^l
\int_0^\infty \e^{-(\mu_1 l + (k-l)/\epsilon)t} dt.
\end{equation}
Performing the integral and afterward taking the limit $\epsilon\to 0$ yields
\begin{equation}
\lim_{\epsilon\to 0} \int_0^\infty \e^{-(\mu_1 l + (k-l)/\epsilon)t} dt = 
\lim_{\epsilon\to 0}\frac{1}{\frac{k-l}{\epsilon}+\mu_1 l} = \frac{1}{\mu_1 l} \delta_{kl},
\label{eq:SM_episonlimit}
\end{equation}
where $ \delta_{kl}$ is the Kronecker delta.
Therefore it follows that
\begin{align}
\langle\tau^+_n\rangle
\leq \frac{1}{\mu_1}\sum_{k=1}^n \binom{n}{k} (-1)^{k+1} \frac{(\mu_1\mtau)^k}{k},
\end{align}
such that we arrive at the inequality
\begin{align}
\langle m_n^+\rangle \equiv 
\langle\tau^+_n\rangle -\mtau
\leq \frac{1}{\mu_1}\sum_{k=1}^n \binom{n}{k} (-1)^{k+1} \frac{(\mu_1\mtau)^k}{k} - \mtau
\equiv \widebar{\mathcal{M}}_n^+,
\end{align}
which completes the proof of the upper bound $\widebar{\mathcal{M}}_n^+$ in Eq.~(14) in the Letter.
\blue{We remark that announced bounds are saturated (i.e., the inequality becomes an equality)
in systems with the above stated spectral gap.}
The validity and sharpness of the upper bound (solid lines) is illustrated
for different model systems and several values of $n$ in Fig.~\ref{fig:SM_Maxdev}.


\subsection{Proof of lower bound on $\langle m_n^+\rangle$}

\begin{figure}[htbp]
\includegraphics[scale=1]{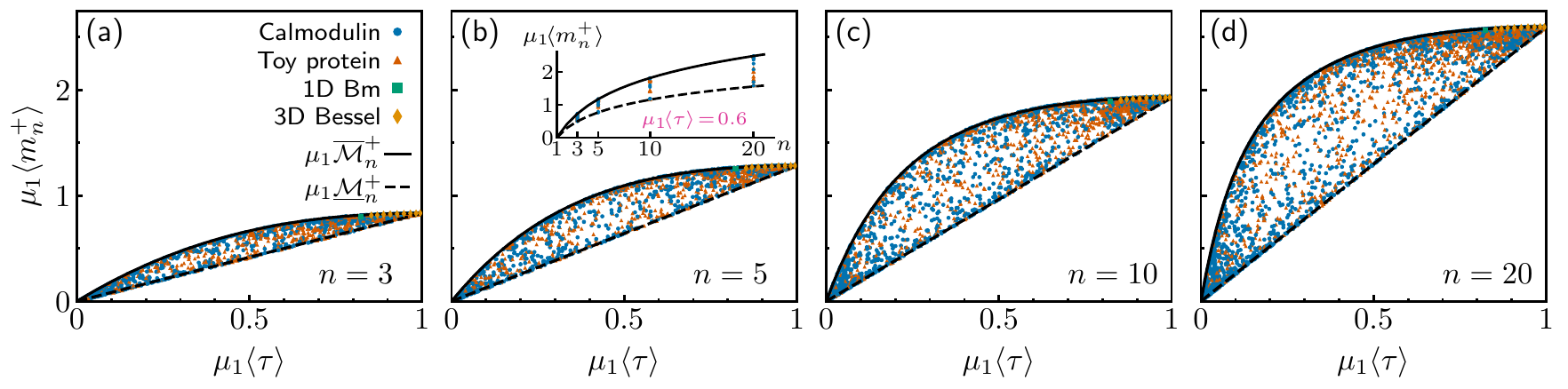}
\caption{{Expected maximal deviation from the mean, $\langle m_n^+\rangle=\langle\tau^+_n\rangle -\mtau$, 
in a sample of (a) $n=3$, (b) $n=5$, (c) $n=10$, and (d) $n=20$ independent realizations
as a function of $\mu_1\mtau$.
Averages are computed over $10^7$ draws of fixed $n$ samples and each data point corresponds to 
a randomly parameterized model of the calmodulin (blue) and toy protein model (red) as discussed in Sec.~\ref{SM_Sampling}. 
Yellow points correspond to the 3D Bessel process with $a\in[0.05,0.95]$ (see Sec.~\ref{SM_SecBessel})
and 1D Brownian motion is shown in green.
Lower bounds (dashed) and upper bounds (solid) are displayed in black, respectively.
(b) The inset shows bounds and numerical data as a function of sample size $n$ for fixed $\mu_1\mtau = 0.6$.}
\label{fig:SM_Maxdev}
}
\end{figure}

To prove a lower bound on 
the expected maximal deviation from the mean 
$\langle m_n^+\rangle$ we first consider the smallest $k$ for which
the corresponding weight $w_k$ is strictly positive, that is, 
$\kplus \equiv {\rm arg\,min}_k w_k > 0$.
The relations $\mtau=\sum_{k\ge 1}w_k/\mu_k$ and $\sum_{k\ge 1}w_k=1$ imply that
\begin{equation}
\frac{w_\kplus}{\mu_\kplus} \leq \mtau \leq \frac{1}{\mu_\kplus},
\label{muk_bound}
\end{equation}
which alongside $w_k\ge 0$ implies the inequalities
\begin{equation}
S(t) \geq w_\kplus \e^{-\mu_\kplus t} \quad\Rightarrow\quad \mathbbm{P}(\tau^+_n\geq t) \geq 1-(1-w_\kplus\e^{-\mu_\kplus t})^n,
\end{equation}
where in the last step we used the relation before Eq.~\eqref{MAXX}.
The integral~\eqref{MAXX} may now be solved recursively. 
To do so we define $\delta_j\equiv\langle\tau^+_j\rangle-
\langle\tau^+_{j-1}\rangle$ such that 
$\langle\tau^+_n\rangle=\sum_{j=1}^n\delta_j$. 
Then we have 
\begin{align}
\delta_j = \int_0^\infty (1-S(t))^{j-1} S(t) dt.
\end{align}
Using the above lower bound, the recursion relation in terms of $\delta_j$ now reads
\begin{equation}
\delta_j \geq \int_0^\infty (1-w_\kplus\e^{-\mu_\kplus t})^{j-1} w_\kplus\e^{-\mu_\kplus t} dt.
\end{equation}
The integral can now be readily solved via the substitution 
$u = 1-w_\kplus \e^{-\mu_\kplus t}$ with $dt = \frac{\e^{-\mu_\kplus t}}{w_\kplus \mu_\kplus}du$
such that we obtain
\begin{equation}
\delta_j \geq \int_0^1 \frac{u^{j-1}}{\mu_\kplus} du  = \frac{1}{j \mu_\kplus}.
\end{equation}
Since $\langle\tau^+_n\rangle = \sum_{j=1}^n \delta_j$ with $\delta_1=\mtau$,
and by Eq.~\eqref{muk_bound} it holds that $1/\mu_\kplus \geq \mtau$,
we obtain the lower bound
\begin{align}
\langle m_n^+\rangle \equiv
\langle\tau^+_n\rangle -\mtau
\geq\frac{1}{\mu_\kplus} \sum_{k=2}^n \frac{1}{k} 
\geq
\mtau \sum_{k=2}^n \frac{1}{k} \equiv \underline{\mathcal{M}}_n^+,
\end{align}
which completes the proof of $\underline{\mathcal{M}}_n^+$ in {Eq.~(14)} of the Letter.
The validity and sharpness of the lower bound (dashed lines) is illustrated
for different model systems and fixed $n$ values in Fig.~\ref{fig:SM_Maxdev}.
\blue{Saturation occurs when $w_{k_+}\to 1$.}

\subsection{Proof of upper bound on $\langle m_n^-\rangle$}
Here, we prove the upper bound on the expected deviation of the minimum 
from the mean in a sample of $n$ i.i.d.~realizations, i.e., 
an upper bound on $\langle m_n^-\rangle$.
According to Eq.~\eqref{eq:SM_MaxMin} we have
\begin{align}
\langle\tau^-_n\rangle 
= \int_0^\infty \mathbbm{P}(\tau^-_n\geq t) dt
= \int_0^\infty S(t)^n dt
= \int_0^\infty \left(\sum_k w_k \e^{-\mu_k t}\right)^n dt
\leq \int_0^\infty \sum_k w_k \e^{-n\mu_k t} dt,
\end{align}
where the last inequality follows by Jensen's inequality since $x^n$ is a convex function
$\forall n\geq 1, x\ge 0$.
After integration of the remaining sum of exponentials we immediately find
\begin{align}
\langle\tau^-_n\rangle 
\leq \int_0^\infty \sum_k w_k \e^{-n\mu_k t} dt
= \sum_k w_k \frac{1}{\mu_k n}
= \frac{\mtau}{n},
\end{align}
since $\mtau = \sum_k\frac{w_k}{\mu_k}$.
Consequently, we arrive at the inequality
\begin{align}
\langle m_n^-\rangle = 
\langle\tau^-_n\rangle - \mtau \leq  \mtau \left[\frac{1}{n}-1\right]\equiv 
\widebar{\mathcal{M}}^-_n,
\label{eq:SM_lowermax}
\end{align}
which proves the upper bound in the second line of Eq.~(14) in the Letter.
\blue{Saturation occurs when $w_k\to 1$ for some $k$ or for degenerate systems with 
identical first-passage eigenvalues $\mu_i = \mu$, $\forall i$ for some $\mu >0$.}
For an illustration of the validity and sharpness of the bound for different model systems
compare Fig.~\ref{fig:SM_Mindev} where $\widebar{\mathcal{M}}^-_n$ is depicted 
as the solid black line.

\subsection{Proof of lower bound on $\langle m_n^-\rangle$}
\label{sec:SM_minlower}
\begin{figure}[htbp]
\includegraphics[scale=1]{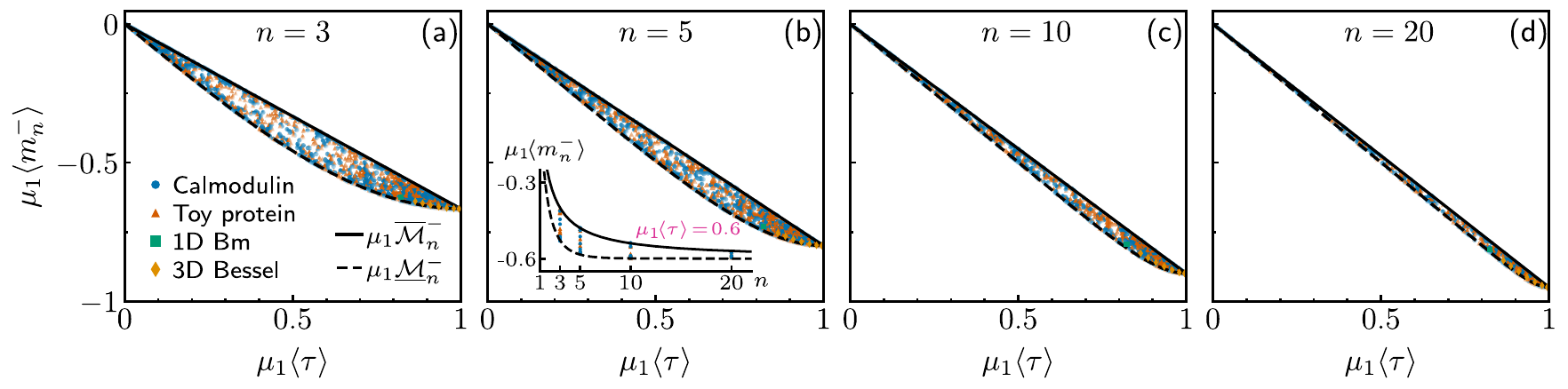}
\caption{{Expected deviation of the minimum from the mean, $\langle m_n^-\rangle=\langle\tau^-_n\rangle -\mtau$, 
in a sample of (a) $n=3$, (b) $n=5$, (c) $n=10$, and (d) $n=20$ independent realizations
as a function of $\mu_1\mtau$.
Averages are computed over $10^7$ draws of fixed $n$ samples and each data point corresponds to 
a randomly parameterized model of the calmodulin (blue) and toy protein model (red) as discussed in Sec.~\ref{SM_Sampling}. 
Yellow points correspond to the 3D Bessel process with $a\in[0.05,0.95]$ (see Sec.~\ref{SM_SecBessel}).
Lower bounds (dashed) and upper bounds (solid) are displayed in black, respectively.
(b) The inset shows bounds and numerical data as a function of
sample-size $n$ for fixed $\mu_1\mtau = 0.6$. }
\label{fig:SM_Mindev}
}
\end{figure}

To prove the lower bound on the expected deviation of the minimum 
from the
mean $\langle\tau^-_n\rangle$ we use an analogous reasoning as in
Sec.~\ref{sec:SM_uppermax}. 
Recall that the ansatz~\eqref{eq:SM_Ansatz} (by the same argument
as in Sec.~\ref{sec:SM_uppermax})  now \emph{minimizes}
$\langle\tau^-_n\rangle$ at fixed $\mu_1\mtau$ and gives 
\begin{equation}
S_{\rm max}(t)^n = \sum_{k=1}^n \binom{n}{k} (1-\mu_1\mtau)^{n-k} \e^{-(n-k)t/\epsilon} (\mu_1\mtau)^k \e^{-\mu_1 k t},
\end{equation}
where we again omit, for the moment, $\lim_{\epsilon\to 0}$ to be
taken later.
Consequently, we have
\begin{equation}
\langle\tau^-_n\rangle 
\geq \int_0^\infty \sum_{k=1}^n \binom{n}{k} (1-\mu_1\mtau)^{n-k} (\mu_1\mtau)^k  \e^{-(\mu_1 k + (n-k)/\epsilon)t} dt.
\end{equation}
Again performing the integral and afterward taking the limit $\epsilon\to 0 $ (compare Eq.~\eqref{eq:SM_episonlimit}) yields 
\begin{equation}
\langle\tau^-_n\rangle \geq
\sum_{k=1}^n \binom{n}{k} (1-\mu_1\mtau)^{n-k} (\mu_1\mtau)^{k} \delta_{nk}
= 
\frac{(\mu_1\mtau)^n}{\mu_1 n},
\end{equation}
and we arrive at the inequality
\begin{equation}
\langle m_n^-\rangle \equiv
\langle\tau^-_n\rangle -\mtau \geq
\frac{(\mu_1\mtau)^n}{\mu_1 n}-\mtau = 
\mtau \left[\frac{(\mu_1\mtau)^{n-1}}{n} -1 \right] 
\equiv \underline{\mathcal{M}}_n^-,
\label{eq:SM_lowermin}
\end{equation}
which completes the proof of the announced lower bound $\underline{\mathcal{M}}_n^-$ in Eq.~(14) in the Letter.
\blue{We note that the bounds are saturated (i.e., the inequality becomes an equality)
in systems with the above stated spectral gap (see Sec.~\ref{sec:SM_uppermax}).}
Finally, we remark that 
the upper bound $\widebar{\mathcal{M}}_n^-$~\eqref{eq:SM_lowermax} and 
lower bound $\underline{\mathcal{M}}_n^-$~\eqref{eq:SM_lowermin} 
approach the value $-\mtau$
from above or below, respectively, as $n\to\infty$, implying that
$\lim_{n\to\infty}\langle\tau^-_n\rangle=0$, as expected.
As can be seen in Fig.~\ref{fig:SM_Mindev}d already for $n=20$ we have
$\widebar{\mathcal{M}}_n^-\to\underline{\mathcal{M}}_n^-\approx -\mtau$.
\newpage
\color{black}
\end{document}